\documentclass[12pt,aps,prd,preprint,tightenlines,
  showpacs,nofootinbib]{revtex4}
\newcommand{\PRE}[1]{{#1}} 

\usepackage{bm}
\usepackage{epsfig}

\newcommand{\mweak}{m_{\text{weak}}}
\newcommand{\mplanck}{m_{\text{Pl}}}

\newcommand{\sigmaan}{\sigma_{\text{an}}}

\newcommand{\OmegaDM}{\Omega_{\text{DM}}}

\newcommand{\kev}{\text{keV}}
\newcommand{\mev}{\text{MeV}}
\newcommand{\gev}{\text{GeV}}
\newcommand{\tev}{\text{TeV}}

\newcommand{\cm}{\text{cm}}
\newcommand{\m}{\text{m}}
\newcommand{\km}{\text{km}}

\newcommand{\s}{\text{s}}

\newcommand{\etal}{{\em et al.}}

\newcommand{\eqref}[1]{Eq.~(\ref{#1})}
\newcommand{\eqsref}[2]{Eqs.~(\ref{#1}) and (\ref{#2})}
\newcommand{\Eqref}[1]{Equation~(\ref{#1})}
\newcommand{\secref}[1]{Sec.~\ref{sec:#1}}
\newcommand{\secsref}[2]{Secs.~\ref{sec:#1} and \ref{sec:#2}}
\newcommand{\Secref}[1]{Section~\ref{sec:#1}}

\newcommand{\figref}[1]{Fig.~\ref{fig:#1}}
\newcommand{\figsref}[2]{Figs.~\ref{fig:#1} and \ref{fig:#2}}

\newcommand{\mphi}{m_{\phi}}
\newcommand{\ephi}{\epsilon_{\phi}}
\newcommand{\vrel}{v_{\text{rel}}}
\newcommand{\vmax}{v_{\text{max}}}
\newcommand{\sigmaanave}{\langle \sigmaan \vrel \rangle}
\newcommand{\sigmaelave}{\langle \sigma_{\text{el}} \vrel \rangle}
\newcommand{\xkd}{x_{\text{kd}}}
\newcommand{\xs}{x_{\text{s}}}
\newcommand{\xnow}{x_{\text{now}}}
\newcommand{\vesc}{v_{\text{esc}}}
\newcommand{\Tkd}{T_{\text{kd}}}

\newcommand{\Tnt}{T_{\text{nt}}}
\newcommand{\alphaEM}{\alpha_{\text{EM}}}
\newcommand{\Seff}{S_{\text{eff}}}
\newcommand{\sv}{(\sigma_{\text{an}} \vrel)_0}

\begin{document}

\preprint{UCI-TR-2010-06}

\title{ \PRE{\vspace*{1.5in}} Sommerfeld Enhancements for Thermal
  Relic Dark Matter
\PRE{\vspace*{0.3in}} }

\author{Jonathan L.~Feng, Manoj Kaplinghat, and Hai-Bo Yu}
\affiliation{Department of Physics and Astronomy, University of
California, Irvine, California 92697, USA \PRE{\vspace*{.5in}} }

\date{August 2010}

\begin{abstract}
\PRE{\vspace*{.3in}} The annihilation cross section of thermal relic
dark matter determines both its relic density and indirect detection
signals.  We determine how large indirect signals may be in scenarios
with Sommerfeld-enhanced annihilation, subject to the constraint that
the dark matter has the correct relic density. This work refines our
previous analysis through detailed treatments of resonant Sommerfeld
enhancement and the effect of Sommerfeld enhancement on freeze out.
Sommerfeld enhancements raise many interesting issues in the freeze
out calculation, and we find that the cutoff of resonant enhancement,
the equilibration of force carriers, the temperature of kinetic
decoupling, and the efficiency of self-interactions for preserving
thermal velocity distributions all play a role.  These effects may
have striking consequences; for example, for resonantly-enhanced
Sommerfeld annihilation, dark matter freezes out but may then
chemically {\em recouple}, implying highly suppressed indirect
signals, in contrast to naive expectations.  In the minimal scenario
with standard astrophysical assumptions, and tuning all parameters to
maximize the signal, we find that, for force-carrier mass $\mphi =
250~\mev$ and dark matter masses $m_X = 0.1$, 0.3, and 1 TeV, the
maximal Sommerfeld enhancement factors are $\Seff = 7$, 30, and 90,
respectively.  Such boosts are too small to explain both the PAMELA
and Fermi excesses.  Non-minimal models may require smaller boosts,
but the bounds on $\Seff$ could also be more stringent, and dedicated
freeze out analyses are required.  For concreteness, we focus on
$4\mu$ final states, but we also discuss $4e$ and other modes,
deviations from standard astrophysical assumptions and non-minimal
particle physics models, and we outline the steps required to
determine if such considerations may lead to a self-consistent
explanation of the PAMELA or Fermi excesses.
\end{abstract}


\maketitle

\section{Introduction}
\label{sec:intro}

Dark matter may be composed of thermal relics with mass near the weak
scale $\mweak \sim 100~\gev - 1~\tev$.  Such candidates are produced
in the hot early Universe and then freeze out when the Universe cools
and expands.  Their annihilation cross section therefore determines
both the relic density and the rate of annihilation today.  The
requirement that the candidate be much or all of the dark matter
therefore constrains its annihilation rate now, with important
implications for indirect searches for dark matter.

Although annihilation in the early Universe and now is determined by
the same dynamics, the kinematics are vastly different: at freeze out,
thermal relics have relative velocity $\vrel \sim 0.3$, whereas today,
$\vrel \sim 10^{-3}$.  As a result, the numerical values of the
annihilation cross sections may differ significantly.  For example,
neutralino annihilation is in many cases dominated by $P$-wave
processes~\cite{Goldberg:1983nd}, and so is suppressed at low
velocities.  In this work, we consider Sommerfeld-enhanced cross
sections, which have the opposite behavior: they are enhanced at low
velocities and therefore boost present-day annihilation signals.
We refine a previous study~\cite{Feng:2009hw} and examine how large
these boosts may be, subject to the constraint that the thermal relic
has the correct thermal relic density to be dark matter.

Along with its general implications for future indirect searches, this
study also has direct implications for the interpretation of current
data. Following earlier excesses reported by the HEAT
Collaboration~\cite{Barwick:1997ig,Beatty:2004cy}, the
PAMELA~\cite{Adriani:2008zr}, ATIC~\cite{:2008zzr}, and
Fermi~\cite{Abdo:2009zk} Collaborations have reported excesses of
cosmic positrons over an estimate of expected
background~\cite{Strong:2009xj}.  Dark matter annihilation with
Sommerfeld enhancement~\cite{Sommerfeld:1931} of dark matter
annihilation~\cite{Baer:1998pg}, generalized to massive force
carriers~\cite{Hisano:2002fk,Hisano:2003ec,%
Hisano:2004ds,Hisano:2005ec,Cirelli:2007xd}, has been proposed as an
explanation~\cite{Cirelli:2008pk,ArkaniHamed:2008qn,%
Pospelov:2008jd,Fox:2008kb}.  In its simplest form, this scenario
assumes that dark matter is composed of a single particle species $X$,
which interacts with light force carriers $\phi$ with fine structure
constant $\alpha_X$ and $\mphi \ll m_X \sim \mweak$.  This new
interaction modifies dark matter annihilation and scattering
properties.  The resulting annihilation cross section multiplied by
relative velocity is $\sv S$, where $\sv$ is its tree-level value, and
$S$ is Sommerfeld's original enhancement factor~\cite{Sommerfeld:1931}
\begin{equation}
S^0 = \frac{\pi \, \alpha_X / v}
{1 - e^{- \pi \alpha_X / v}} \ \
\stackrel{\alpha_X \gg v}{\longrightarrow} \ \
\frac{\pi \alpha_X}{v} \ ,
\label{S0}
\end{equation}
generalized to massive $\phi$, where $v \equiv \vrel/2$ is the dark
matter particle's velocity in the center-of-mass frame.  In the
proposed explanation, dark matter freezes out at early times when the
Sommerfeld effect is negligible with $\sv \approx 3\times
10^{-26}~\cm^3/\s$, leading to the correct relic density $\OmegaDM h^2
\simeq 0.114$.  At present, however, when $\vrel$ is much smaller, the
Sommerfeld effect becomes important, and, for example, for $m_X \sim
2~\tev$ and an assumed enhancement factor of $S \sim 1000$, such
annihilations are sufficient to explain the positron
excesses~\cite{Bergstrom:2009fa,Meade:2009iu}.

In Ref.~\cite{Feng:2009hw}, we showed that in straightforward models,
such large Sommerfeld enhancements cannot be self-consistently
realized.  The problem is simple to state: large Sommerfeld
enhancement requires strong interactions, and strongly-interacting
particles annihilate too efficiently in the early Universe to be all
of the dark matter.  More quantitatively, the required new force
carrier interaction necessarily induces an annihilation process $XX
\to \phi \phi$, with cross section
\begin{equation}
\sv \sim \frac{\pi \alpha_X^2}{m_X^2} \ .
\label{sigma}
\end{equation}
Conservatively neglecting other annihilation processes, for a typical
weak-scale mass $m_X$, the thermal relic density implies a typical
weak coupling $\alpha_X \approx 0.05 \, [m_X / 2~\tev]$, as dictated
by the WIMP miracle.  The resulting Sommerfeld enhancement now is
$\sim \pi \alpha_X / v\approx 100 \, [m_X / 2~\tev]$, an order of
magnitude too small to explain the PAMELA and Fermi signals.
Alternatively, to achieve $S \sim 1000$, one requires $\alpha_X \sim
1$, which implies $\Omega_X \sim 0.001$, two orders of magnitude
smaller than the value assumed in deriving the requirement $S \sim
1000$.

In this work, we refine our previous analysis in several ways.  In
Ref.~\cite{Feng:2009hw} we approximated the Sommerfeld enhancement $S$
by its value at $\mphi = 0$, given by $S^0$ in \eqref{S0}.  For
massive $\phi$, the Sommerfeld enhancement cuts off at a value
proportional to $m_X/\mphi$ and also exhibits resonant structure.
Here we use numerical results (and a highly accurate analytic
approximation to the numerical
results~\cite{Iengo:2009ni,Iengo:2009xf,Cassel:2009wt,Slatyer:2009vg})
for $S$.

In addition, we refine our previous work to include the effect of
Sommerfeld enhancement on freeze out.  One might expect this effect to
be negligible, because freeze out is typically thought to occur at $v
\sim 0.3$, when the Sommerfeld effect is insignificant.  However,
annihilation continues to much later times, and Sommerfeld enhancement
has an impact when the dark matter cools.  Sommerfeld effects on
freeze out have been considered previously in
Ref.~\cite{Hisano:2006nn} for the case of Wino dark matter and in
Refs.~\cite{MarchRussell:2008yu,Yuan:2009bb,Dent:2009bv,Zavala:2009mi}
for hidden sector dark matter.  In the analyses published after the
PAMELA and Fermi excesses were
reported~\cite{Yuan:2009bb,Dent:2009bv,Zavala:2009mi}, $\sv$ was taken
as a free parameter, and the fact that it depends on the same
parameters that determine $S$ was not used.  Here we make essential
use of the observation that there is a irreducible contribution to
$\sv$ of the form of \eqref{sigma}, and so constraints on the relic
density bound the fundamental parameters that also determine $S$.

As we will see, the combination of these two refinements highlights a
number of effects that may typically be ignored, but now require
exploration.  For example, the relic density is in principle affected
by the cutoff of resonant Sommerfeld enhancement, the production and
decay rates of force carrier particles, the temperature of kinetic
decoupling, and the efficiency of self-scattering for preserving the
dark matter's thermal velocity distribution.  Very close to
resonances, we also find the intriguing possibility of chemical {\em
recoupling}, in which dark matter freezes out and then melts back in,
with annihilations becoming important again at late times.  This
implies that exact resonances suppress, rather than enhance, indirect
signals.

We examine the quantitative impact of all of these effects, and
present results for the maximal Sommerfeld enhancement of indirect
signals.  In the most optimistic case, that is, tuning all parameters
to maximize the signal, we find that, for $\mphi = 250~\mev$ and $m_X
= 0.1$, 0.3, and 1 TeV, the largest possible effective Sommerfeld
enhancements are $\Seff = 7$, 30, and 90, respectively. This refined
analysis therefore strengthens our previous results: Sommerfeld
enhancements may be significant, but they fall short of explaining the
current PAMELA and Fermi cosmic ray excesses.  We then discuss various
astrophysical effects that may reduce the discrepancy between the
maximal enhancements derived here and those required to explain the
data, and critically examine several non-minimal models and the issues
that must be addressed to determine if more complicated particle
physics models may provide viable explanations.

This paper is organized as follows.  In \secref{S}, we describe our
underlying model assumptions and our treatment of Sommerfeld
enhancements with resonances.  In \secref{freezeout} we analyze the
effect of Sommerfeld enhancement on freeze out.  In \secref{results}
we assemble these pieces and present the results for the maximal
Sommerfeld enhancement achievable now.  In \secref{comparison}, we
compare these to the current data, and discuss non-standard
astrophysical effects and non-minimal particle physics models.  We
present our conclusions in \secref{conclusions}.

\section{Sommerfeld Enhancement with Resonances}
\label{sec:S}

We consider a simple model with a hidden sector dark matter particle
$X$, which couples to a light force carrier $\phi$ with coupling
$\sqrt{4 \pi \alpha_X}$.  The annihilation cross section is then $\sv
S$, where $\sv$ is the tree-level cross section and $S$ is the
Sommerfeld enhancement.

To maximize the Sommerfeld enhancement, we assume that $\sv$ is
dominated by $S$-wave processes, and so is unsuppressed at low
velocities.  We also consider only the ``irreducible'' annihilation
channel $X X \to \phi \phi$, and take the tree-level cross section
\begin{equation}
\sv = \frac{\pi \alpha_X^2}{m_X^2} \ .
\label{sigma0}
\end{equation}
This may be modified by ${\cal O}(1)$ pre-factors, depending, for
example, on whether $X$ is a Majorana or Dirac fermion, and whether
$\phi$ is a scalar or a gauge boson.  In addition, even in simple
models, $\sv$ will typically receive additional contributions from
other annihilation channels; to maximize the Sommerfeld effect on
indirect search signals, we neglect these other channels here, but
discuss their impact in \secref{comparison}.

To determine the enhancement factor $S$, we numerically solve the
differential equation
\begin{equation}
\frac{1}{m_X} \frac{d^2\chi}{dr^2} + \frac{\alpha_X}{r} e^{-\mphi r}
\chi = - m_X v^2 \chi \ ,
\end{equation}
with the boundary conditions $\chi'(r)=im_X v \chi(r)$ and
$\chi(r) = e^{im_X v r}$ when $r \to \infty$. The Sommerfeld
enhancement factor is given by
\begin{equation}
S = \frac{\left| \chi( \infty ) \right|^2}
{\left| \chi(0) \right|^2} \ .
\label{Snumerical}
\end{equation}

\begin{figure}[tb]
\begin{center}
\includegraphics*[width=0.49\columnwidth]{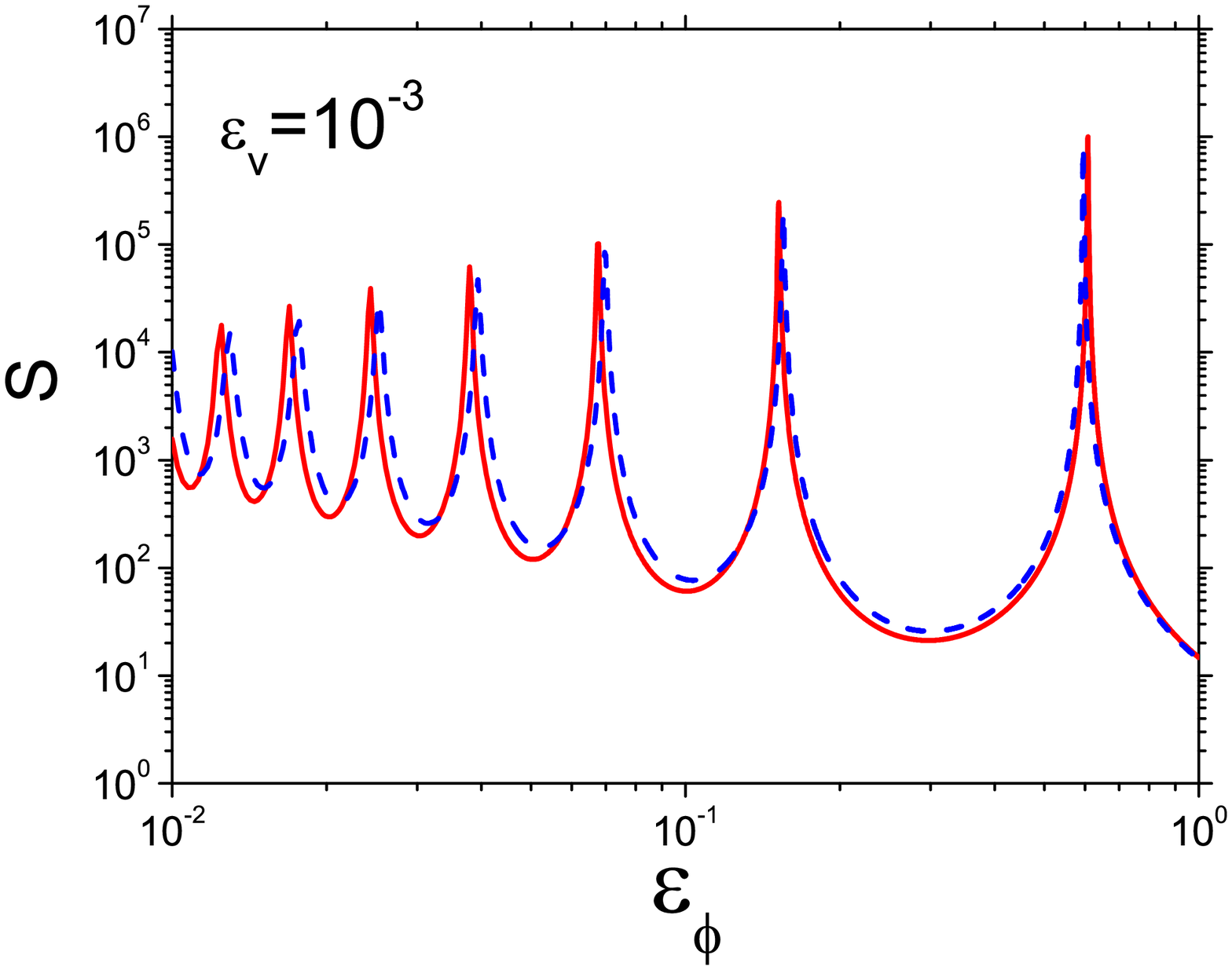}
\includegraphics*[width=0.49\columnwidth]{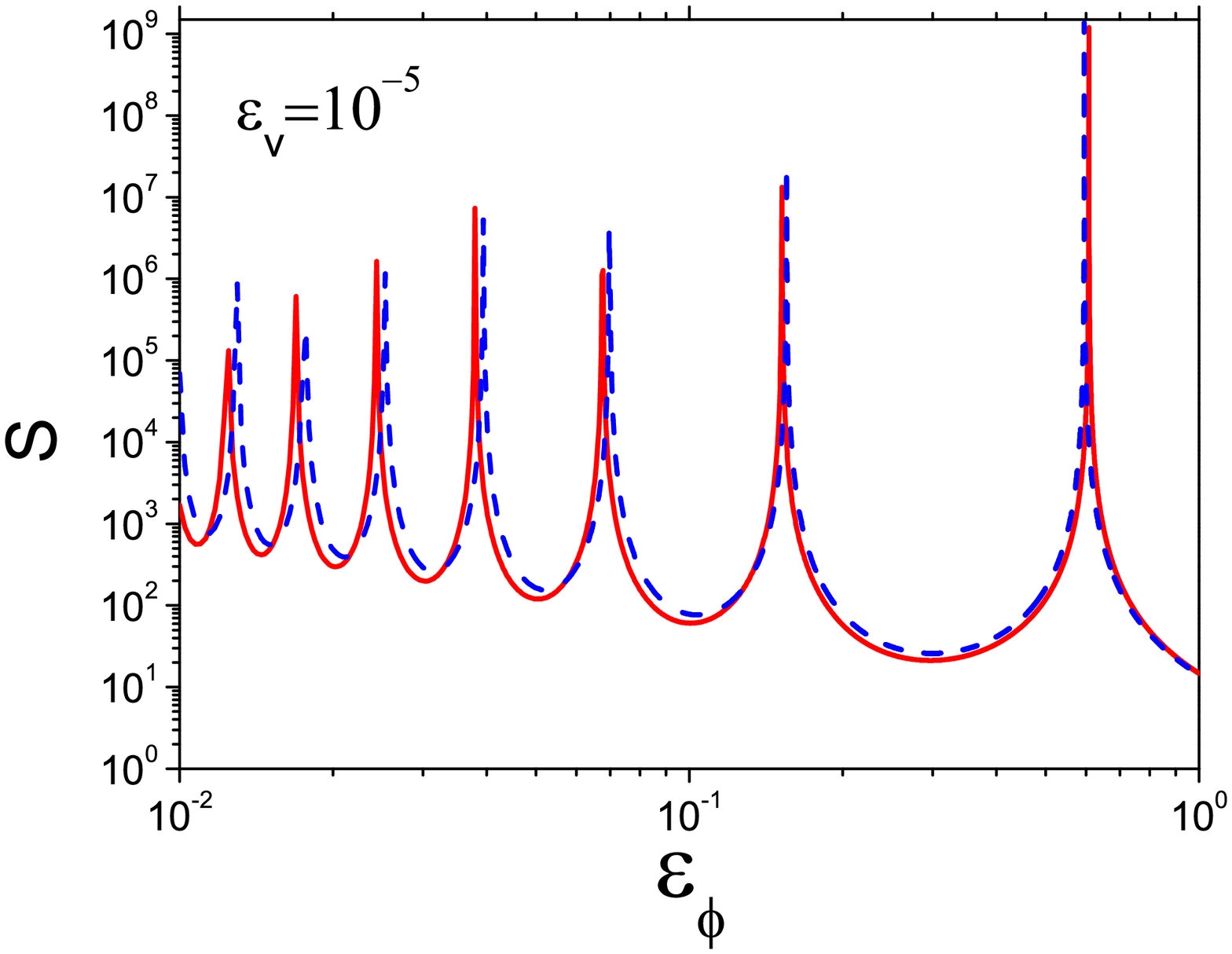}
\end{center}
\vspace*{-.25in}
\caption{The Sommerfeld enhancement factor $S$ as a function of $\ephi
\equiv \mphi/ (\alpha_X m_X)$ for the constant values of $\epsilon_v
\equiv v/\alpha_X$ indicated.  The solid red curves are the analytic
approximation of \eqref{Sapprox}, and the dashed blue curves are
numerical results.
\label{fig:Sfactor}}
\end{figure}

The Sommerfeld enhancement may also be obtained by approximating the
Yukawa potential by the Hulthen potential, for which an analytic
solution is possible~\cite{Cassel:2009wt}.  The resulting analytic
approximation to the Sommerfeld enhancement
is~\cite{Cassel:2009wt,Slatyer:2009vg}
\begin{equation}
S = \frac{\pi}{\epsilon_v}
\frac{\sinh \left( \frac{2\pi\epsilon_v}{\pi^2\ephi/6}
  \right)}
{ \cosh \left( \frac{2\pi\epsilon_v} {\pi^2 \ephi / 6} \right)
- \cos \left( 2\pi\sqrt{\frac{1}{\pi^2\ephi/6}
-\frac{\epsilon^2_v}{(\pi^2\ephi/6)^2}} \, \right) } \ ,
\label{Sapprox}
\end{equation}
where $\epsilon_v \equiv v / \alpha_X$ and $\ephi \equiv \mphi/
(\alpha_X m_X)$.  The analytic expression of \eqref{Sapprox} is
compared to the numerical solution in \figref{Sfactor}.  We see that
the analytic result is an excellent approximation, typically
reproducing the numerical results to within fractional differences of
10\%, and accurately reproducing the resonant behavior.  Given
\eqref{Sapprox}, we see that for $\epsilon_v \ll \epsilon_{\phi}$,
these resonances are at
\begin{equation}
\mphi \simeq \frac{6 \alpha_X m_X}{\pi^2 n^2} \ , \quad
n = 1, 2, 3 \ldots
\label{mphiresonance}
\end{equation}
At these resonances, with $\mphi$ determined by \eqref{mphiresonance},
the Sommerfeld enhancement for low $v$ is
\begin{equation}
S \simeq \frac{\pi^2 \alpha_X \mphi}{6 m_X v^2} \ .
\end{equation}
$S$ is therefore enhanced by $v^{-2}$ at resonances, as
opposed to $v^{-1}$ away from resonances.

Note, however, that both the numerical and analytic results become
infinite on resonance.  This is unphysical, a result of the fact that
the quantum mechanical treatments do not include the effect of bound
state decay.  In fact, the finite lifetime of bound states implies
that $S$ saturates at $v \sim \alpha_X^3
(\mphi/m_X)$~\cite{Hisano:2004ds}, which, given the condition for a
strong resonance, \eqref{mphiresonance} with small $n$, is $v \sim
\alpha_X^4$.  In this study, we use the analytic form for $S$ given in
\eqref{Sapprox} and include the saturation by making the substitution
$v \to v + \alpha_X^4$.

\section{Thermal Freeze Out with Sommerfeld Enhancement}
\label{sec:freezeout}

\subsection{General Formalism}
\label{sec:formalism}

Given values of $m_X$, $\mphi$, and $\alpha_X$, we must determine the
relic density $\Omega_X$.  The formalism of thermal freeze out is well
developed~\cite{Gondolo:1990dk,Kolb:1990vq}.  Here we review this
formalism in sufficient generality to accommodate novel effects
resulting from Sommerfeld enhancements.

The evolution of the abundance of a thermal relic $X$ is governed by
the Boltzmann equation
\begin{equation}
\frac{d n_X}{dt} + 3 H n_X = - \sigmaanave
\left( n_X^2 - n_X^{\text{eq}\, 2} \right) ,
\end{equation}
where $n_X$ is the number density of the dark matter particles,
$n_X^{\text{eq}}$ is its value in equilibrium, $H$ is the Hubble
constant, and $\sigmaanave$ is the annihilation cross section
multiplied by the relative velocity, averaged over the dark matter
velocity distribution. Changing variables from $t \to x=m_X/T$ and
$n_X \to Y=n_X/s$, where $s$ is the entropy density, the Boltzmann
equation becomes
\begin{equation}
\frac{dY}{dx} = - \sqrt{\frac{\pi}{45}} \, \mplanck \, m_X
\frac{( g_{*s} / \sqrt{g_*} \, )}{x^2}
\sigmaanave \left( Y^2 - Y^{\text{eq}\, 2} \right) ,
\label{YBoltzmann}
\end{equation}
where $\mplanck = G_N^{-1/2} \simeq 1.2 \times 10^{19}~\gev$ is the
Planck mass, and $g_{*s}$ and $g_*$ are the effective relativistic
degrees of freedom for entropy and energy density, respectively.  In
this section we will assume that the particles $X$ annihilates to are
in thermal equilibrium at the time of freeze out. We will revisit this
issue in \secref{phidecay}, where we show that this requirement leads
to significant constraints if the dominant annihilation is to the dark
force carriers.

To evaluate the relic density, we first evolve it to the time of
chemical decoupling, when the annihilation rate $\Gamma_{\text{an}}
\equiv n_X \sigmaanave$ is approximately the expansion rate $H$.  As
discussed in \secref{S}, we assume the annihilation cross section has
the form $\sigma_{\text{an}} \vrel = \sv S$, where $\sv$ is the
$S$-wave, tree-level cross section, and $S$ is the Sommerfeld
enhancement.  Up to the time of freeze out, the Sommerfeld effect is
insignificant, and so the standard results for $S$-wave annihilators
apply: defining freeze out by
\begin{equation}
Y(x_f) = (1 + c) \, Y^{\text{eq}}(x_f) \ ,
\label{yxf}
\end{equation}
where $c$ is a constant, freeze out occurs at
\begin{eqnarray}
x_f & \approx & \ln \xi - \frac{1}{2} \ln \left( \ln \xi \right)
\nonumber \\
\xi & = & 0.038 \, c \, (2+c) \, \mplanck \, m_X
\left( g / \sqrt{g_*} \, \right) \sv \ ,
\label{xf}
\end{eqnarray}
where $g$ is the number of degrees of freedom of the dark matter
particle; we take $g=2$.  To match numerical results, $c \sim 1$; we
choose $c=1$ and have checked that varying $c$ from 1/2 to 2 has no
appreciable effect.  To further test this approximation we have also
numerically evolved \eqref{YBoltzmann} from $x=20$ to $x=100$ (when
the equilibrium abundance is effectively zero) and then used
\eqref{xf} to evolve into the late Universe. This numerical
calculation typically results in 10\% smaller relic abundance.  Given
the excellent agreement, we set $c=1$ and with this choice,
\eqsref{yxf}{xf} determine $Y(x_f)$, the abundance at freeze out.

After freeze out, $Y^{\text{eq}}$ quickly becomes insignificant.
Neglecting it, we may solve \eqref{YBoltzmann} to find
\begin{eqnarray}
\frac{1}{Y(\xs)} &=& \frac{1}{Y(x_f)}
+ \sqrt{\frac{\pi}{45}} \mplanck \, m_X  \int_{x_f}^{\xkd}
\frac{(g_{*s}/\sqrt{g_*}) \sigmaanave}{x^2} dx \nonumber \\
&& \qquad \quad + \sqrt{\frac{\pi}{45}} \mplanck \, m_X  \int_{\xkd}^{\xs}
\frac{(g_{*s}/\sqrt{g_*}) \sigmaanave}{x^2} dx \ ,
\label{Yfinal}
\end{eqnarray}
where $\xkd$ is the value of $x$ at kinetic decoupling and $\xs$ is
its value when annihilations become insignificant and we may stop the
integration.  We have broken the integral into two to emphasize that
there are two eras: before and after the temperature of kinetic
decoupling $\Tkd$.  Before $\Tkd$, the dark matter's velocity
distribution is thermal with temperature $T_X = T$. After $\Tkd$, the
dark matter's velocity distribution initially remains thermal, but
$T_X$ drops as $a^{-2}$, while $T$ drops as $a^{-1}$, where $a$ is the
scale factor, and so $T_X = T^2/\Tkd$. Eventually, the dark matter's
velocity distribution need not even be thermal.  We discuss the value
of $\Tkd$ and the issue of non-thermal velocity distributions in
\secsref{kd}{velocity}, respectively.

When the dark matter distribution is thermal with temperature $T_X$,
the thermally-averaged cross section in the non-relativistic limit is
\begin{eqnarray}
\sigmaanave &\equiv& \int f(\vec{v}_1) f(\vec{v}_2) \sigmaan \vrel
d^3 \vec{v}_1 d^3 \vec{v}_2 \nonumber \\
&=& \int \sqrt{\frac{2}{\pi}} \frac{1}{v^3_0} \vrel^2
e^{-\frac{\vrel^2}{2v^2_0}} \sigmaan \vrel d\vrel \ ,
\end{eqnarray}
where $f(\vec{v})$ is the dark matter's velocity distribution, $\vrel
= |\vec{v}_1 - \vec{v}_2|$, and
\begin{equation}
v_0 = \sqrt{\frac{2 T_X}{m_X}} \equiv \sqrt{\frac{2}{x_X}} \ ,
\end{equation}
the most probable velocity.  For the Sommerfeld-enhanced annihilation
cross section, this becomes
\begin{equation}
\sigmaanave =\frac{x^{3/2}_X}{2\sqrt{\pi}}\int^{\infty}_0(\sigmaan
\vrel) \vrel^2 e^{-x_X \vrel^2 / 4} d \vrel =\sv {\bar S}(x_X) \ ,
\end{equation}
where
\begin{equation}
{\bar S}(x_X) = \frac{x^{3/2}_X}{2\sqrt{\pi}} \int_0^{\infty}
S \vrel^2 e^{-x_X \vrel^2 /4} d\vrel
\label{Sbar}
\end{equation}
is the Sommerfeld enhancement averaged over a thermal distribution
with temperature $T_X = m_X/x_X$.

\subsection{Equilibration of Force Carriers}
\label{sec:phidecay}

In the relic density calculation, we have implicitly assumed that,
when the $X$ particles chemically decouple, the force carriers $\phi$
are in thermal equilibrium with the massless particles of the standard
model.  The force carriers are expected to interact with the standard
model, as their decays to standard model particles typically provide
the indirect signals.  In a simple example, if $\phi$ is a U(1) gauge
boson, it may mix with the standard model photon through kinetic
mixing terms $\sim \epsilon F_{\mu \nu}^{\text{EM}} F^{\phi \,
\mu\nu}$.  After diagonalizing the $\phi$-photon system, standard
model particles with charge $Q$ have hidden charge $\epsilon
Q$~\cite{Holdom:1985ag}, and so the $\phi$ particles decay through
$\phi \to f \bar{f}$, where $f = e, \mu, u, d, s, \ldots$. The largest
kinetic mixing parameter allowed by current particle physics
constraints is $\epsilon \sim
10^{-3}$~\cite{Batell:2009yf,Bjorken:2009mm}.

The existence of $\phi$ interactions with the visible sector does not,
however, guarantee that they are efficient enough to bring the $\phi$
particles in thermal equilibrium.  Here we determine sufficient
conditions for the kinetic mixing example to guarantee the
equilibration of $\phi$ particles.  The 
leading $\phi$ number-changing interactions between the $\phi$
particles and the visible sector are decays $\phi \to f\bar{f}$ and
inverse decays $f \bar{f} \to \phi$.  Other $\phi$ number-changing
processes, such as $f \bar{f} \to \phi \gamma$, $q \bar{q} \to \phi
g$, and $f \gamma \to f \phi$, are parametrically suppressed compared
to these and subdominant.  For temperatures $T \gg \mphi$, the decay
and inverse decay processes balance (since $f,\bar{f}$ are in chemical
equilibrium), and we may simply compare the decay rate to the
expansion rate ~\cite{Pospelov:2007mp}. 

For $T \gg \mphi$, the thermally-averaged decay rate for $\phi \to f
\bar{f}$ in the lab frame is $\langle \Gamma_\phi \rangle \simeq
(\epsilon^2/3) \sum_f (Q_f^2 N^f_c) \alphaEM \mphi (\mphi/T)$, where
we have averaged over three $\phi$ polarizations, $Q_f$ and $N^f_c$
are the standard model charge and number of colors for fermion $f$,
the sum is over all fermions with mass $m_f < \mphi / 2$, and the last
factor of $\mphi/T$ is from time dilation. Note that the decay process
is inefficient at early times and high temperatures, but as $T$ drops,
$\langle \Gamma_\phi \rangle$ increases and $H$ decreases, and so the
$\phi$ particles may come into thermal equilibrium with the the
standard model.  The expansion rate is $H \simeq 1.66 \, g_{*}^{1/2}
T^2 / \mplanck$; at freezeout, $g_{*} \simeq 100$.  Requiring $\langle
\Gamma_\phi \rangle \agt H$ at $T_f \simeq m_X /25$ yields $\epsilon
\agt 3 \times 10^{-6}$ for $\mphi=\gev$ and $m_X=\tev$.

If the above constraint is not satisfied, then the hidden sector may
be at a different temperature from the standard model, modifying the
dark matter freezeout.  Such scenarios have been considered in
Ref.~\cite{Feng:2008mu}, and the difference in temperatures has a
significant impact on the dark matter relic density.  These
considerations may be important for some models, but here we assume
that \eqref{phiequilibrium} is satisfied and $\mphi \ll T_f \simeq
m_X/25$, so that the $\phi$ particles are in thermal contact with the
standard model at freeze out and their number density is close to the
thermal prediction.

We note here that the calculation above does not guarantee chemical
equilibrium. For the $\phi$ particles to produce $X$ through $\phi\phi
\rightarrow XX$, we need the tail of the $\phi$ distribution to be
populated thermally. To check this, we approximate the decay rate for
$\phi$ with energies of order $m_X$ as $(\epsilon^2/3) \sum_f (Q_f^2
N^f_c) \alphaEM \mphi (\mphi/m_X)$.  If this decay rate is larger than
the expansion rate, then the inverse process will also be in
equilibrium and hence thermally populate the tail of the $\phi$
distribution.  Comparing this to the expansion rate at $T_f \simeq
m_X/25$ we obtain that
\begin{equation}
\epsilon \agt 1.5 \times 10^{-5}
\left[ \frac{4}{\sum_f Q_f^2 N^f_c} \right]^{\frac{1}{2}}
\left[ \frac{\gev}{\mphi} \right]
\left[ \frac{m_X}{\tev} \right]^{\frac{3}{2}} \ .
\label{phiequilibrium}
\end{equation}
In setting the bound above, we have not considered scattering
interactions with standard model fermions (and the resulting changes
in kinetic energy for $\phi$) because these processes are proportional
to $\epsilon^4$ and hence considerably slower that the inverse decay
process.

To summarize, for the $\phi$ particles to be in thermal equilibrium
with the standard model at freezeout requires that
\eqref{phiequilibrium} is satisfied and $\mphi \ll T_f \simeq m_X/25$.
We will assume these conditions hold in our analysis, but we note that
they do not necessarily hold.  In particular, if $\mphi \sim T_f$, the
number of $\phi$ particles at freezeout is reduced, which reduces the
$X$ relic density, strengthening the bounds on $\Seff$ we determine
below.

\subsection{Kinetic Decoupling of Dark Matter}
\label{sec:kd}

After freeze out, dark matter particles remain kinematically coupled
to the thermal bath through elastic scattering.  Kinetic decoupling
occurs later, when the momentum transfer rate drops below the Hubble
expansion rate.  We define the kinetic decoupling temperature $\Tkd$
by $\Gamma_k(\Tkd)=H(\Tkd)$, where $\Gamma_k$ is the momentum transfer
rate.  It may be approximated as
\begin{equation}
\Gamma_k \sim  n_r \sigmaelave \frac{T}{m_X} \ ,
\end{equation}
where $n_r$ is the number density of the relativistic species in the
thermal bath, and $\sigmaelave$ is the thermally-averaged cross
section for elastic scattering between dark matter particles and the
relativistic particles in the thermal bath.

We first consider the model-independent elastic scattering off the
hidden sector thermal bath of $\phi$
particles~\cite{Finkbeiner:2009mi}. For $T \alt \mphi$, $n_{\phi}$ is
Boltzmann suppressed, and so this process will not be able to maintain
kinetic equilibrium. For $T \agt \mphi$, the thermally-averaged $X
\phi \to X \phi$ cross section is
\begin{equation}
\sigmaelave \sim \frac{\alpha^2_X}{m_X^2} \ ,
\end{equation}
the Thomson scattering cross section.  Given the $\phi$ number density
$n_{\phi} \sim T^3$, the momentum transfer rate is $\Gamma_k \sim
\alpha_X^2 T^4/m_X^3$, and this is equal to $H \sim T^2 / \mplanck$ at
\begin{equation}
T \sim \left[ \frac{m_X^3}{\alpha^2_X \mplanck} \right]
^{\frac{1}{2}}
= 0.43~\mev
\left[ \frac{0.021}{\alpha_X} \right]
\left[ \frac{m_X}{\tev} \right]^{\frac{3}{2}} \ ,
\end{equation}
where we have normalized the expression to typical parameters that
give the correct relic density in the presence of Sommerfeld
enhancement.  Halo shape constraints require $\mphi \agt
30~\mev$~\cite{Feng:2009hw}, and so for all physically viable and
relevant parameters,
\begin{equation}
\Tkd^\phi \sim \mphi \ .
\label{Tkdphi}
\end{equation}
This is the temperature of kinetic decoupling from the hidden thermal
bath.  It is quite model-independent, as it assumes only the
$X$-$\phi$ interactions required in all Sommerfeld enhancement
scenarios, and so it provides a maximal value of $\Tkd$ in Sommerfeld
scenarios.

Dark matter particles may also be kept in kinetic equilibrium through
scattering off the visible sector's thermal bath.  This is more model
dependent, but as discussed in \secref{phidecay}, if $\phi$ is a U(1)
gauge boson, it may mix with the standard model photon.  After
diagonalizing the $\phi$-photon system, standard model particles with
charge $Q$ have hidden charge $\epsilon Q$, inducing new energy
transfer processes. The most efficient process is $Xe \to Xe$
scattering through $t$-channel $\phi$ exchange.  For $T \alt m_e$,
Boltzmann suppression of the electron number density makes the
interaction inefficient.  At temperatures $\mphi \agt T \agt m_e$, the
corresponding cross section is~\cite{Feng:2009mn}
\begin{equation}
 \sigmaelave \sim
\frac{ \epsilon^2 \alphaEM \alpha_X T^2}{\mphi^4} \ .
\end{equation}
The momentum transfer rate $\Gamma_k \sim \epsilon^2 \alphaEM \alpha_X
T^6/(\mphi^4 m_X)$ is equal to the expansion rate at temperature
\begin{equation}
T \sim \left[ \frac{\mphi^4 m_X}{\epsilon^2 \alphaEM
    \alpha_X \mplanck }\right]^{1/4} \ ,
\end{equation}
and so the resulting temperature of kinetic decoupling from the
visible sector's thermal bath is
\begin{equation}
\Tkd^e (\epsilon) \sim \max \left\{ m_e \, , 0.82~\mev
\left[ \frac{10^{-3}}{\epsilon} \right]^{\frac{1}{2}}
\left[ \frac{\mphi}{30~\mev} \right]
\left[ \frac{0.021}{\alpha_X} \right]^{\frac{1}{4}}
\left[ \frac{m_X}{\tev} \right]^{\frac{1}{4}} \right\} \ ,
\label{Tkde}
\end{equation}
where we have again normalized $\alpha_X$ and $m_X$ to typical freeze
out parameters, and additionally normalized $\mphi$ to its smallest
possible value and $\epsilon$ to its largest allowed value.  When
$\epsilon$ is near its maximal value, $\Tkd^e (\epsilon) <
\Tkd^{\phi}$, and so interactions with the visible thermal bath delay
kinetic decoupling.  We denote the lowest possible kinetic decoupling
temperature
\begin{equation}
\Tkd^e \equiv \Tkd^e (\epsilon = 10^{-3}) \ ,
\end{equation}
and will explore the dependence of our results on the temperature of
kinetic decoupling by varying it between its maximal value
$\Tkd^{\phi}$ and its minimal value $\Tkd^e$.

By crossing symmetry, the visible sector scattering interaction also
implies an annihilation process $XX \to e^+ e^-$.  In principle, this
also enters the thermal relic density calculation.  The cross section
is $\sigmaanave \sim \epsilon^2 \alphaEM \alpha_X / m_X^2$, however,
and the $\epsilon^2$ suppression makes this subdominant for all
relevant cases.

\subsection{Velocity Distribution after Kinetic Decoupling}
\label{sec:velocity}

Before kinetic decoupling, dark matter particles have the same
temperature as the thermal bath and a Maxwell-Boltzmann phase space
distribution.  Usually one assumes that the phase space distribution
remains Maxwell-Boltzmann after kinetic decoupling.  However, this is
not necessarily true in scenarios with Sommerfeld-enhanced
annihilation, because slow particles annihilate with larger cross
sections.  This preferentially depletes the low velocity population
and may distort the phase space distribution. In this case, the
standard formulae used to compute the relic density of dark matter are
not valid and one needs to explicitly consider the effect of
annihilations using the full phase space distribution.

As discussed in Refs.~\cite{Ackerman:2008gi,Feng:2009mn,%
Feng:2009hw,Buckley:2009in,Ibe:2009mk}, however, the $\phi$ field
mediates self-scattering between dark matter particles. If the
self-scattering rate is higher than the Hubble expansion rate, the
momentum exchanged in these self-interactions with be sufficient to
maintain thermal equilibrium.  The momentum transfer of particles
interacting through Yukawa potentials has been studied in
Refs.~\cite{Khrapak:2003,Khrapak:2004}.  Although the authors of these
studies were interested in slow and highly charged particles moving in
plasmas with screened Coulomb potentials, they approximated these
potentials by Yukawa potentials, and so their results are exactly
applicable in the current particle physics context.  The numerical
results of these studies for the momentum transfer cross section are
accurately reproduced by~\cite{Khrapak:2003,Khrapak:2004}
\begin{eqnarray}
\sigma_T &\approx& \frac{4\pi}{m^2_\phi} \beta^2 \ln(1+\beta^{-1}) \ ,
\quad \beta<0.1 \nonumber \\
\sigma_T &\approx& \frac{8\pi}{m^2_\phi}
\frac{\beta^2}{1+1.5\beta^{1.65}} \ ,
\quad 0.1 \le \beta \le 1000 \\
\sigma_T &\approx& \frac{\pi}{m^2_\phi}
\left( \ln\beta + 1 - \frac{1}{2} \ln^{-1} \beta \right)^2  ,
\quad \beta > 1000 \ , \nonumber
\end{eqnarray}
where $\beta = 2 \alpha_X \mphi / (m_X \vrel^2)$.

\begin{figure}[tb]
\begin{center}
\includegraphics*[width=0.49\columnwidth]{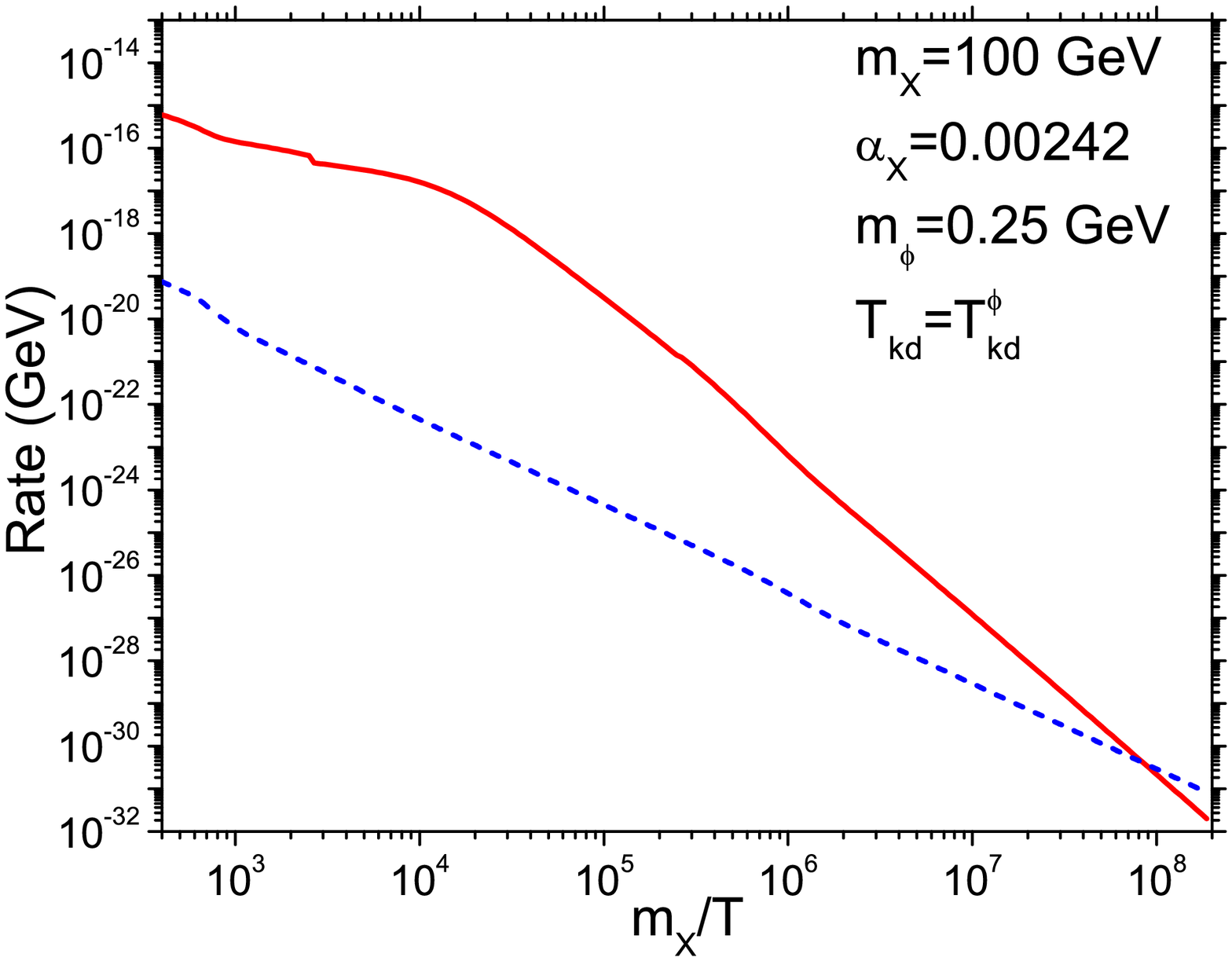}
\includegraphics*[width=0.49\columnwidth]{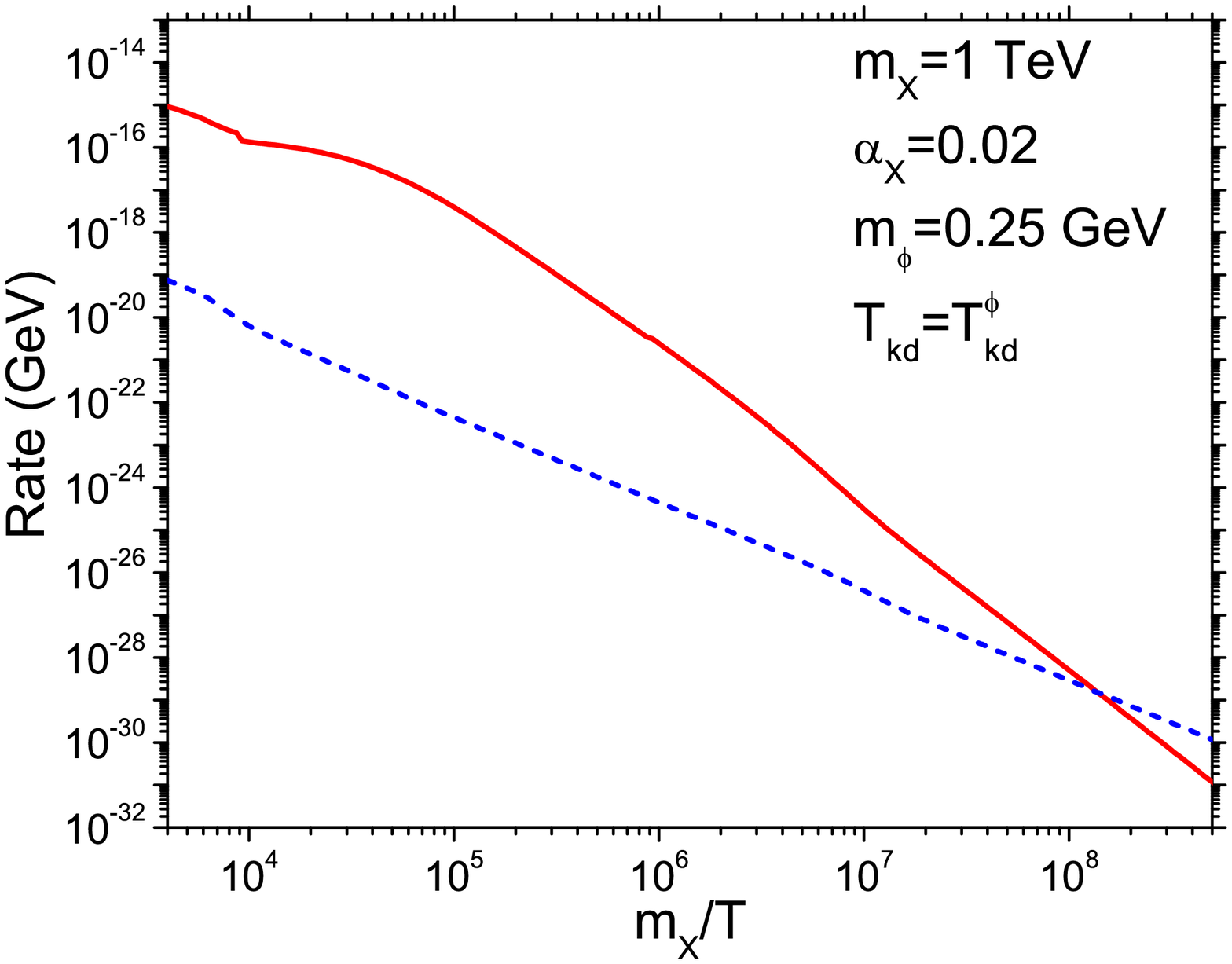}
\end{center}
\vspace*{-.25in}
\caption{The self-scattering rate $\Gamma_s$ (solid red) and the
Hubble rate $H$ (dashed blue) as functions of $x = m_X/T$ for the
values of $m_X$, $\mphi$, $\alpha_X$, and $\Tkd$ indicated.
\label{fig:scatteringrate}}
\end{figure}

Given these results, we may calculate the dark matter self-scattering
rate and compare it to the Hubble expansion rate.  The self-scattering
rate is given by~\cite{Feng:2009hw}
\begin{equation}
\Gamma_s = \int f(\vec{v}_1) f(\vec{v}_2) \, n_X \left(\sigma_T \vrel
\right) \frac{\vrel^2}{v_0^2} d^3\vec{v}_1 d^3\vec{v}_2 \ ,
\end{equation}
where $n_X$ is the dark matter density. This sets the time scale to
change velocities by ${\cal O}(1)$. To find the rate to change the
kinetic energy by ${\cal O}(1)$, one would divide the above rate by a
factor of 3. However, such details are not important for the present
calculation. $H$ and $\Gamma_s$ are presented in
\figref{scatteringrate} for two representative cases.

We may define the self-scattering decoupling temperature $\Tnt$ by
$\Gamma_s(\Tnt)=H(\Tnt)$; after $\Tnt$, the dark matter velocity
distribution may become non-thermal.  To derive an approximate
expression for $\Tnt$, note that to have the right relic abundance,
the co-moving number density of dark matter during the self-scattering
epoch cannot be smaller than the co-moving number density at present.
The dark matter number density $n_X$ may therefore be taken to be
\begin{equation}
n_X \sim s \left(\frac{n_X}{s}\right)_0
= s \, \frac{\OmegaDM \rho_c}{m_X s_0}
\simeq s \times 3.9 \times 10^{-13} \left[ \frac{\tev}{m_X} \right] \ .
\end{equation}
During the self-scattering decoupling epoch, typically $\beta > 1000$,
and so $\sigma_T \sim \kappa/\mphi^2$, where $\kappa \sim \pi \ln^2
\beta \sim 600-1000$ for the parameters of interest.  With this
approximate expression for $\sigma_T$, we find
\begin{equation}
\Tnt \sim 20~\kev
\left[ \frac{\mphi}{250~\mev} \right]
\left[ \frac{m_X}{1~\tev} \right]^{\frac{3}{4}}
\left[ \frac{\Tkd}{250~\mev} \right]^{\frac{1}{4}}
\left[ \frac{\kappa}{800} \right]^{-\frac{1}{2}} \ .
\end{equation}
This agrees well with our numerical results, and we use this
approximate form with $\kappa = 800$ in deriving our results below.

{}From both the numerical and analytical results, we see that the
self-scattering rate efficiently preserves the thermal distribution to
temperatures $\Tnt \sim 10~\kev$, when dark matter velocities are
$\vrel \sim T/(\Tkd m_X)^{1/2} \sim 10^{-5}$.  This is a low velocity,
and in most cases, we will find that it is sufficiently low that the
impact of non-thermality after $\Tnt$ has a negligible impact on the
thermal relic density.  Note, however, that $\Tnt$ is not low enough to
ensure a thermal distribution down to $\vrel \sim \alpha_X^4$, where
the Sommerfeld resonances cut off, and so very close to resonances,
the non-thermality will have an effect.

\section{Results}
\label{sec:results}

\subsection{Maximal Sommerfeld Enhancements}
\label{sec:maximal}

In this section, we present results for the maximal Sommerfeld
enhancement.  We have defined several Sommerfeld factors, including
$S^0$, the original Sommerfeld factor for massless $\phi$ given in
\eqref{S0}, and $S$, the Sommerfeld factor generalized to massive
$\phi$, which includes resonances and is given in
\eqsref{Snumerical}{Sapprox}.  Here we also define the effective
Sommerfeld enhancement factor
\begin{equation}
\Seff = \frac{\sv \bar{S}_{\text{now}}} {3\times10^{-26}~\cm^3/\s} \ ,
\end{equation}
where
\begin{equation}
\bar{S}_{\text{now}} \simeq \frac{\xnow^{3/2}}{2 \sqrt{\pi} N}
\int_0^{\vmax} S \vrel^2 e^{-\xnow \vrel^2 /4} d\vrel \ ,
\label{Snowapprox}
\end{equation}
$\xnow \equiv 2/v_0^2$, and $N = \text{erf} \, ( z / \sqrt{2} \, ) -
(2/\pi)^{1/2} z e^{-z^2/2}$ with $z \equiv \vmax / v_0$. $\Seff$ is the
experimentally relevant parameter, as it is the factor by which
indirect fluxes are enhanced relative to the case without Sommerfeld
enhancement.  Unlike the early Universe case, in the halo we have to
cut off the velocity integral at some maximum relative speed $\vmax$,
which is related to the escape speed from the local neighborhood. This
maximum speed is a function of the angle between the dark matter
particles in the lab frame. However, we have checked that setting
$\vmax$ to be equal to the escape speed $\vesc$ and integrating as in
\eqref{Snowapprox} makes a difference of less than 10\%. The escape
speed at about 10 kpc from the center of the halo is estimated to be
about $500~\km/\s$ to $550~\km/\s$~\cite{Xue:2008se}. We use $\vesc =
525~\km/\s$ to derive our results, noting that 10\% variations in
$\vesc$ have a much smaller effect on $\Seff$. We have assumed that
the velocity dispersion tensor is isotropic with the 1-D velocity
dispersion given by $v_0/\sqrt{2}$. We use $v_0 = 210~\km/\s$
consistent with determinations of the circular velocity
(e.g.,~\cite{Xue:2008se}). Inferring the dispersion from the circular
velocity curve requires knowledge of the dark matter density
profile. The value we use for $v_0$ is consistent with a
Navarro-Frenk-White density profile with a scale radius of about 20
kpc. Here again, we note that there are uncertainties in the value of
$v_0$ at the 10\% level. This translates directely into an uncertainty
in $\Seff$ of about 10\% if $S \propto 1/v$ and 20\% around resonance
where $S\propto 1/v^2$.

For comparison purposes, we also define
\begin{equation}
\Seff^0 = \frac{\sv \bar{S}_{\text{now}}^0}
{3\times10^{-26}~\cm^3/\s} \ ,
\end{equation}
where
\begin{equation}
\bar{S}_{\text{now}}^0 = \frac{\xnow^{3/2}}{2 \sqrt{\pi} N}
\int_0^{\vesc} S^0 \vrel^2 e^{-\xnow \vrel^2 /4} d\vrel
\end{equation}
is the effective Sommerfeld enhancement without resonances, but with
the Sommerfeld effect on freeze out included.

\begin{figure}[tb]
\begin{center}
\includegraphics*[width=0.49\columnwidth]{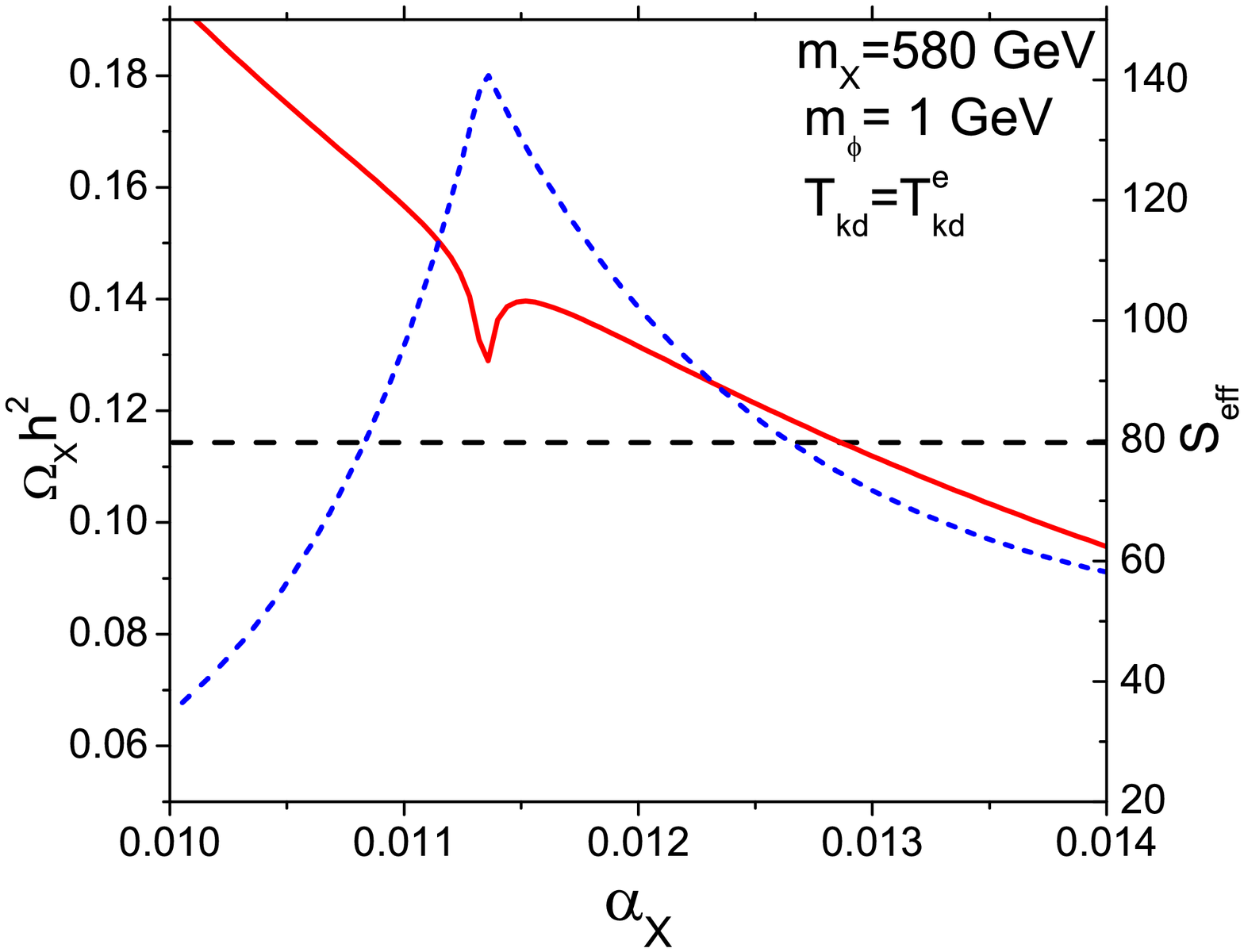}
\includegraphics*[width=0.49\columnwidth]{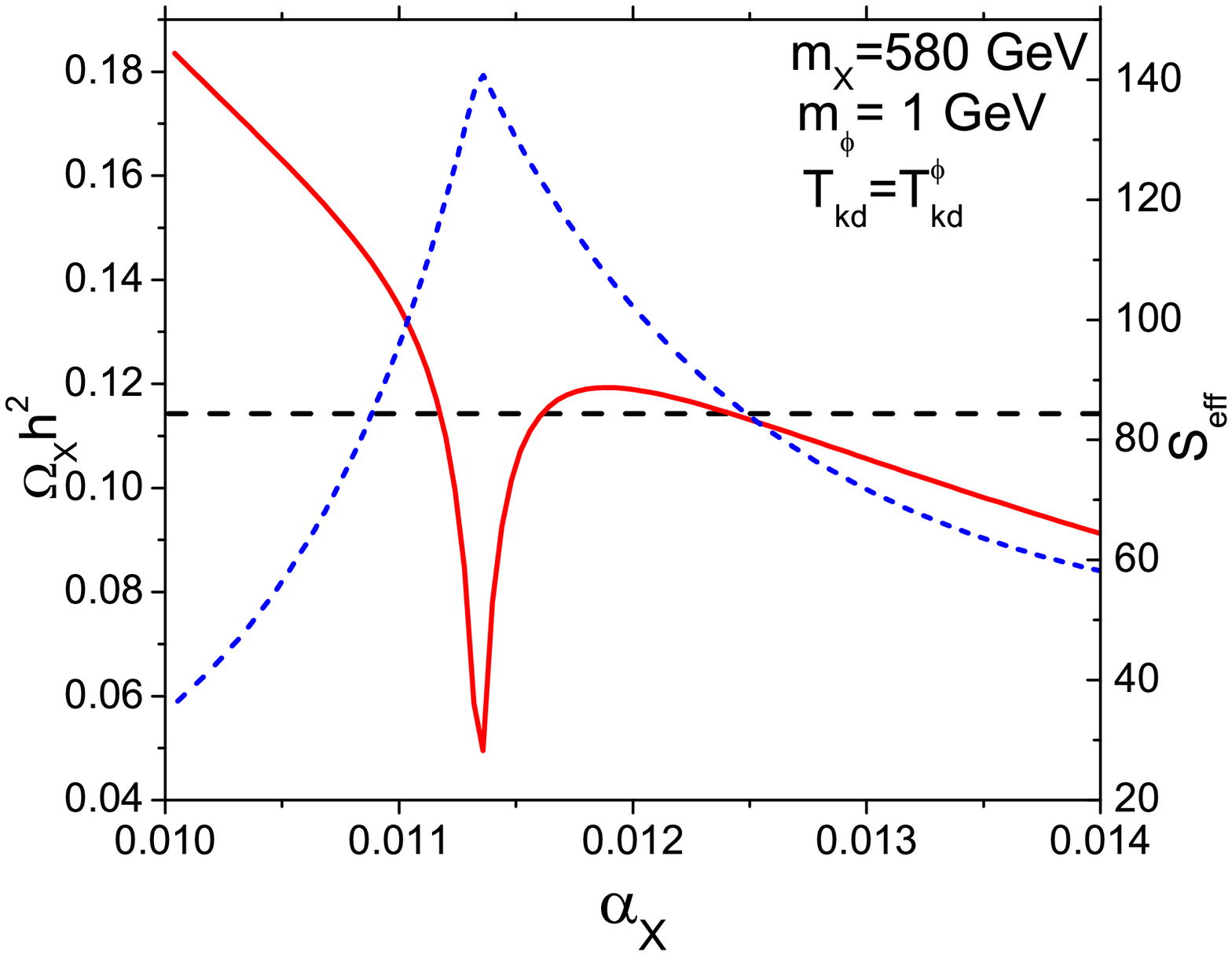}
\end{center}
\vspace*{-.25in}
\caption{The relic density $\Omega_X h^2$ (solid red) and $\Seff$
(dotted blue) as a function of the fine-structure constant $\alpha_X$
for the fixed values of $m_X$, $\mphi$, and $\Tkd$ indicated and the
observed value of $\Omega_X h^2$ (dashed black).  In most cases, there
is a unique choice of $\alpha_X$ that yields the correct $\Omega_X h^2
= 0.114$ (left).  In the presence of strong resonances, however,
there are cases where three different choices of $\alpha_X$ all yield
the correct $\Omega_X$ (right). In these cases, $\Seff$ varies by
about 20--30\% between the different solutions.
\label{fig:volcano}}
\end{figure}

\begin{figure}[tb]
\begin{center}
\includegraphics*[width=0.49\columnwidth]{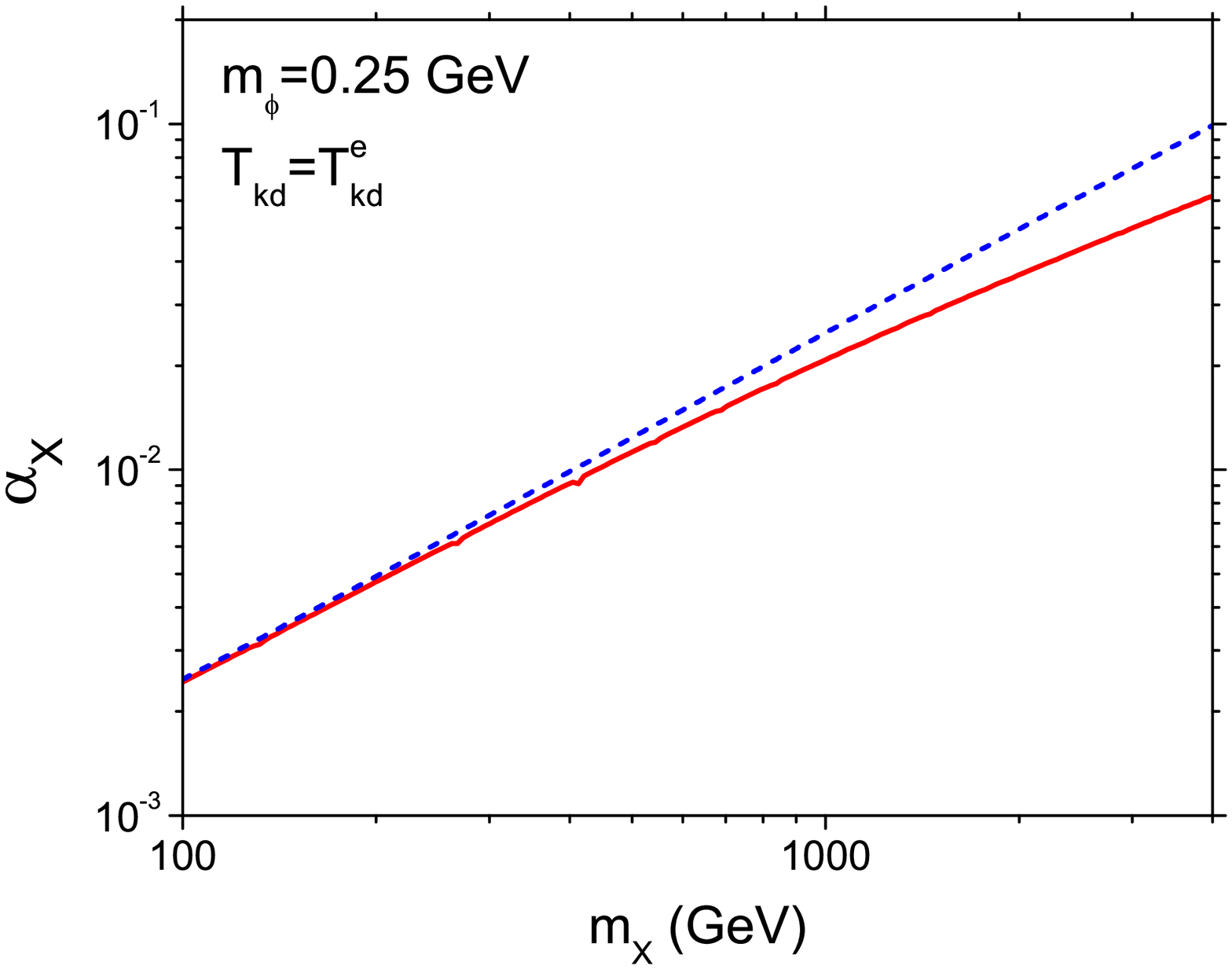}
\includegraphics*[width=0.49\columnwidth]{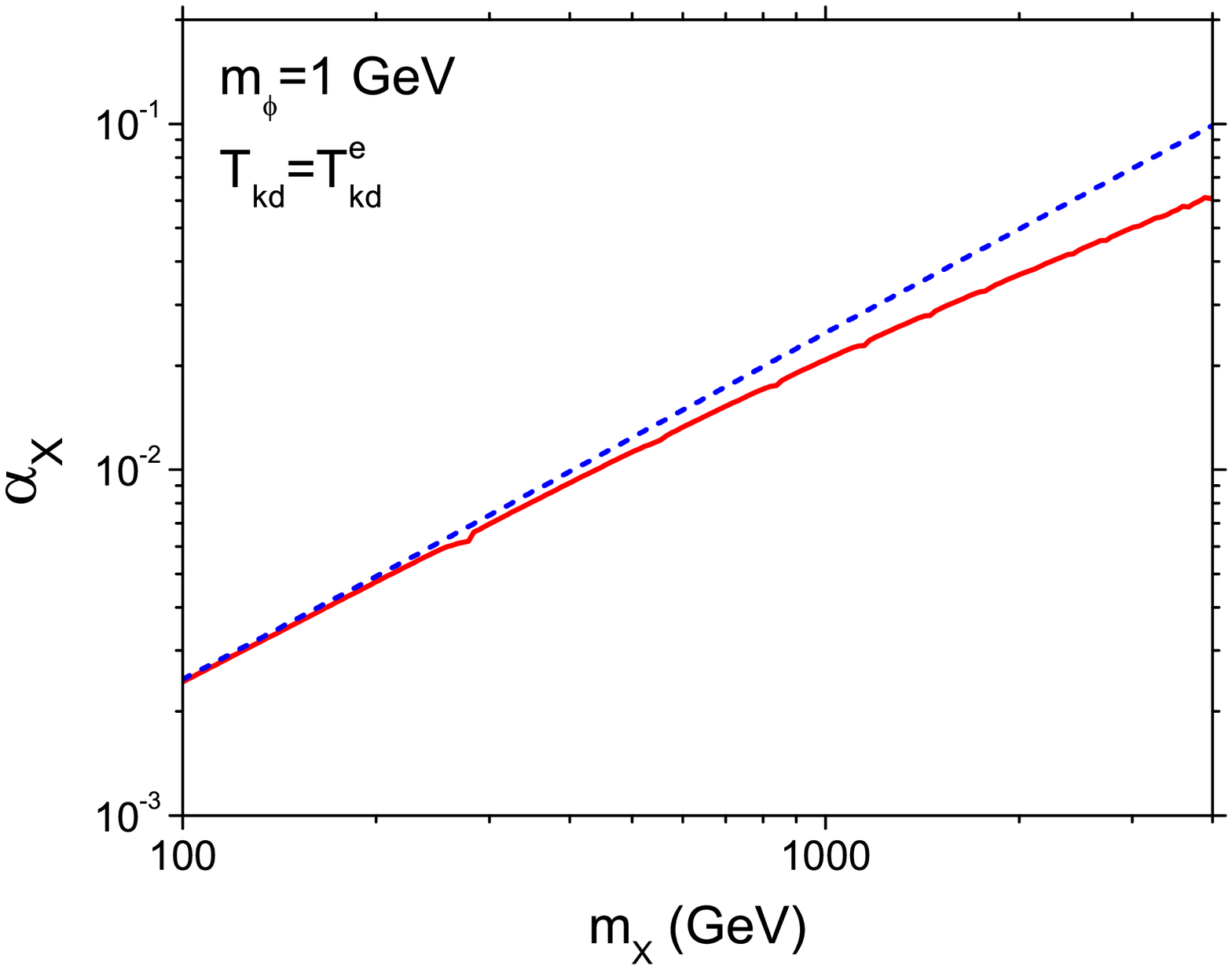}
\end{center}
\vspace*{-.25in}
\caption{The value of $\alpha_X$ required to achieve a relic density
  of $\Omega_X h^2=0.114$ as a function of the dark matter particle
  mass $m_X$ (solid red) for $\mphi=250~\mev$ (left) and
  $\mphi=1~\gev$ (right). We also plot the required $\alpha_X$ (dotted
  blue) if Sommerfeld effects are neglected in the early Universe. The
  tree level cross section (without Sommerfeld enhancement) is
  $\sv=\pi \alpha_X^2/m_X^2$. Because $\alpha_X$ varies over almost
  two orders of magnitude in this plot, the dips near resonance are
  not immediately apparent.
\label{fig:alphamx}}
\end{figure}

To calculate the largest possible $\Seff$ for a given $m_X$ and
$\mphi$, we make a number of optimistic ($\Seff$-maximizing)
assumptions:
\begin{itemize}
\setlength{\itemsep}{1pt}\setlength{\parskip}{0pt}\setlength{\parsep}{0pt}
\item We fix $\Tkd = \Tkd^e$.  This delays kinetic decoupling as much
  as possible, keeps the dark matter as hot as possible, and so
  reduces the Sommerfeld effect on freeze out, maximizing $\Seff$.
\item We fix $\xs$ in \eqref{Yfinal} by stopping the dark matter evolution
  at $\Tnt$; that is, we neglect all annihilations that occur after
  the dark matter distribution becomes non-thermal.  This is certainly
  optimistic, as the distribution will remain thermal for some time
  and dark matter annihilations will continue, but this again minimizes
  the Sommerfeld effect on freeze out, maximizing $\Seff$.
\item We require $\Omega_X h^2 = 0.114$.  This might appear to be too
  restrictive; after all, there is no requirement that the observed
  signals arise from a particle that makes up all of the dark matter.
  However, if a flux arises from Sommerfeld-enhanced annihilation, it
  scales as $n_X^2 \sigmaanave S \sim \alpha_X^{-1}$, because its number
  density scales as $n_X \sim \Omega_X \sim \sigmaanave^{-1} \sim
  \alpha_X^{-2}$ and $S \sim \alpha_X$.  The flux can therefore always
  be increased by lowering $\alpha_X$ until $\Omega_X h^2$ is all of
  the dark matter, and so choosing $\Omega_X h^2$ in fact maximizes
  indirect signals.
\item We choose the maximal $\alpha_X$ that yields $\Omega_X h^2 =
  0.114$.  Roughly speaking, $\Omega_X h^2$ decreases as $\alpha_X$
  increases, and so typically, there is a unique choice of $\alpha_X$
  that yields the correct $\Omega_X h^2$.  In some cases with strong
  resonances, however, there are three choices of $\alpha_X$ that give
  the correct $\Omega_X$, as shown in \figref{volcano}.  When we do a
  coarse scan there is no guarantee that we always pick the solution
  with the largest value of $\Seff$.  Note, however, that the values
  of $\Seff$ only change by about 20--30\% between the three allowed
  solutions, and such variations do not modify our conclusions. The
  values of $\alpha_X$ that yield the correct relic density for given
  values of $m_X$ and $\mphi$ are given in \figref{alphamx}.  Note
  that close to a resonance, one needs extreme fine-tuning to avoid
  efficient annihilation in the early Universe and obtain the correct
  relic density.
\end{itemize}

\begin{figure}[tb]
\begin{center}
\includegraphics*[width=0.49\columnwidth]{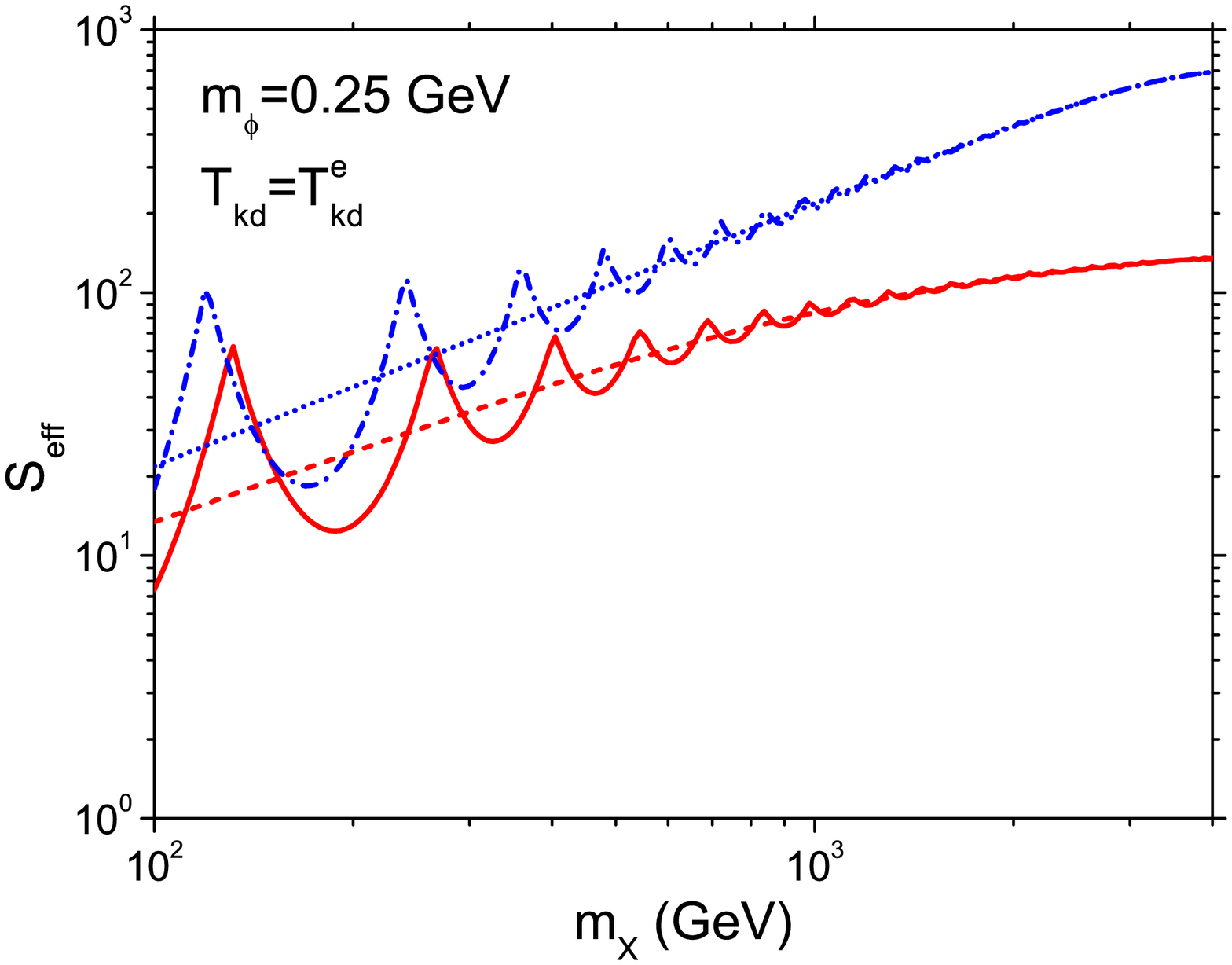}
\includegraphics*[width=0.49\columnwidth]{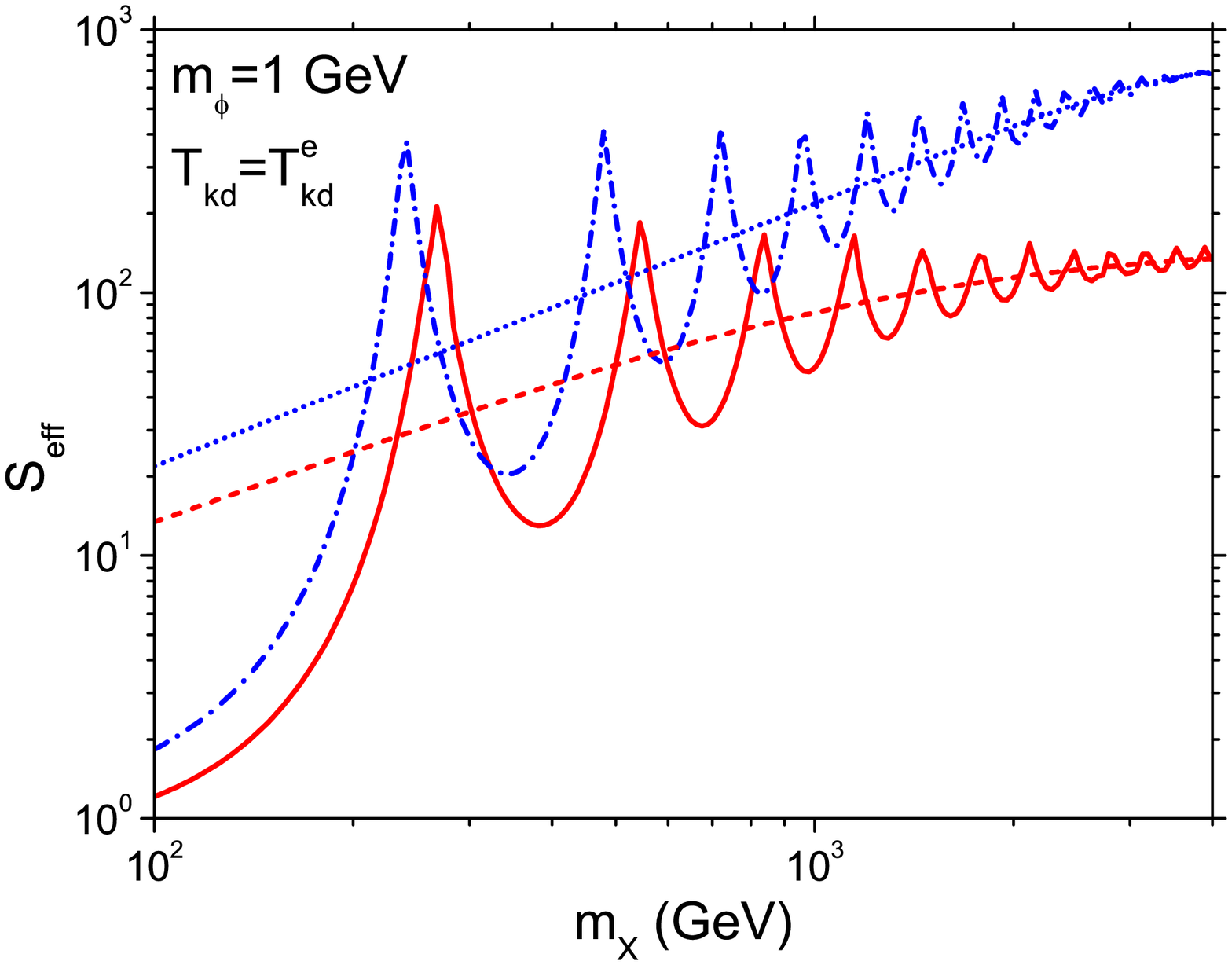}
\end{center}
\vspace*{-.25in}
\caption{The effective Sommerfeld enhancement factor $\Seff$ (solid
red) as a function of $m_X$ for $\mphi = 250~\mev$ (left) and 1 GeV
(right) and the set of $\Seff$-maximizing assumptions listed in
\secref{maximal}.  Also shown for comparison are $S^0$ (dotted blue),
the Sommerfeld factor without resonances and neglecting the Sommerfeld
effect on freeze out; $\bar{S}$ (dot-dashed blue), the Sommerfeld
factor with resonances but neglecting the Sommerfeld effect on freeze
out; and $\Seff^0$ (dashed red), the Sommerfeld factor without
resonances, but including the Sommerfeld effect on freeze out.
\label{fig:Seff_mx}}
\end{figure}

In \figref{Seff_mx}, we show the maximal values of $\Seff$ as a
function of $m_X$ for $\mphi = 250~\mev$ and 1 GeV.  To understand the
impact of resonances and the Sommerfeld effect on freeze out on these
results, we also plot other Sommerfeld enhancements.  $S^0$, which
includes neither resonances nor freeze out effects, was used in
Ref.~\cite{Feng:2009hw}.  We see that it is almost always
overestimates the maximal Sommerfeld enhancement.  This is because, as
evident in \figref{Seff_mx}, the effect of Sommerfeld enhanced
annihilation on freeze out is a significant suppression.  This may be
understood as follows. If Sommerfeld enhancement reduces the thermal
relic density by a factor $\zeta$, the tree-level cross section $\sv$
must be reduced by $\zeta$ to keep $\Omega_X$ fixed.  However, $\Seff
\propto \alpha_X^3 \propto \sv^{3/2}$, and so reducing $\Omega_X$ by
$\zeta$ implies a reduction in $\Seff$ by a factor $\zeta^{3/2}$.  For
example, for $m_X = 1~\tev$, the maximal coupling is $\alpha_X \simeq
0.021$, and $\zeta \sim 1.5$, consistent with the results of
Refs.~\cite{Dent:2009bv,Zavala:2009mi}.  Including the effect of
freeze out here therefore reduces $S^0$ to $\Seff^0$ by a factor of
$(1.5)^{3/2} \sim 2$, a significant reduction.

Adding resonances then produces oscillations about the $S^0$ and
$\Seff^0$ contours, with peak positions given by
\eqref{mphiresonance}.  We see that, although resonances can
significantly enhance the effective Sommerfeld factor, these
enhancements are significant only for low values of $n$; for $\mphi
\alt \gev$, these lie at low $m_X$, and the effect becomes negligible
for $m_X \agt 1~\tev$.

\begin{figure}[tb]
\begin{center}
\includegraphics*[width=0.69\columnwidth]{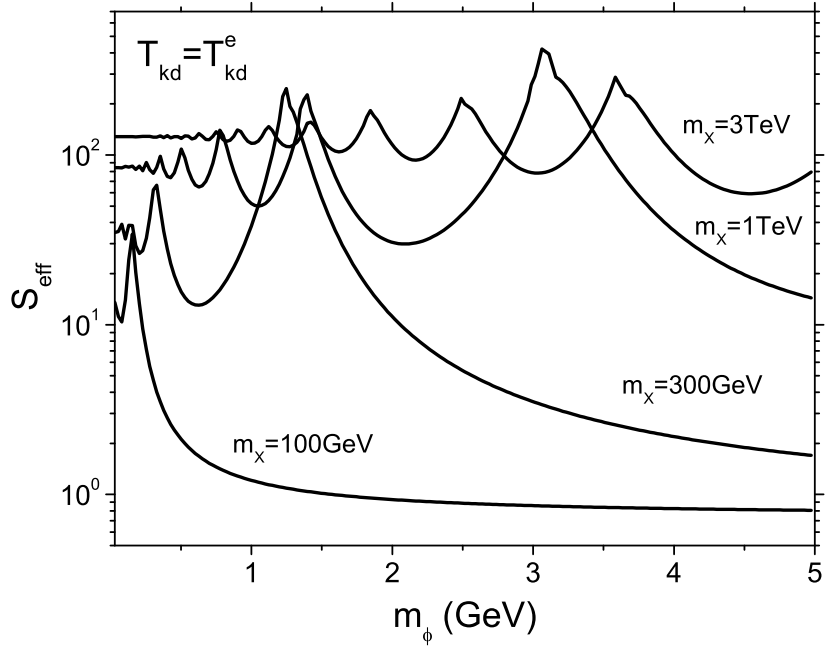}
\end{center}
\vspace*{-.25in}
\caption{The effective Sommerfeld enhancement factor $\Seff$ as a
function of $\mphi$ for the $m_X$ indicated and the set of
$\Seff$-maximizing assumptions listed in \secref{maximal}.
\label{fig:Seff_mphi}}
\end{figure}

In \figref{Seff_mx}, we have fixed $\mphi$ to typical values
considered in the literature.  In \figref{Seff_mphi}, we plot upper
bounds on $\Seff$ as a function of $\mphi$ for fixed values of $m_X$.
We see that the resonances have little impact for $\mphi \alt 1~\gev$,
but can produce enhancements by factors of 2 to 3 for larger $\mphi$.
Large values of $\mphi \agt 1~\gev$ have been considered disfavored,
however, as they eliminate the kinematic suppression of anti-proton
fluxes, which are typically considered to be consistent with
astrophysical backgrounds.

\begin{figure}[tb]
\begin{center}
\includegraphics*[width=0.49\columnwidth]{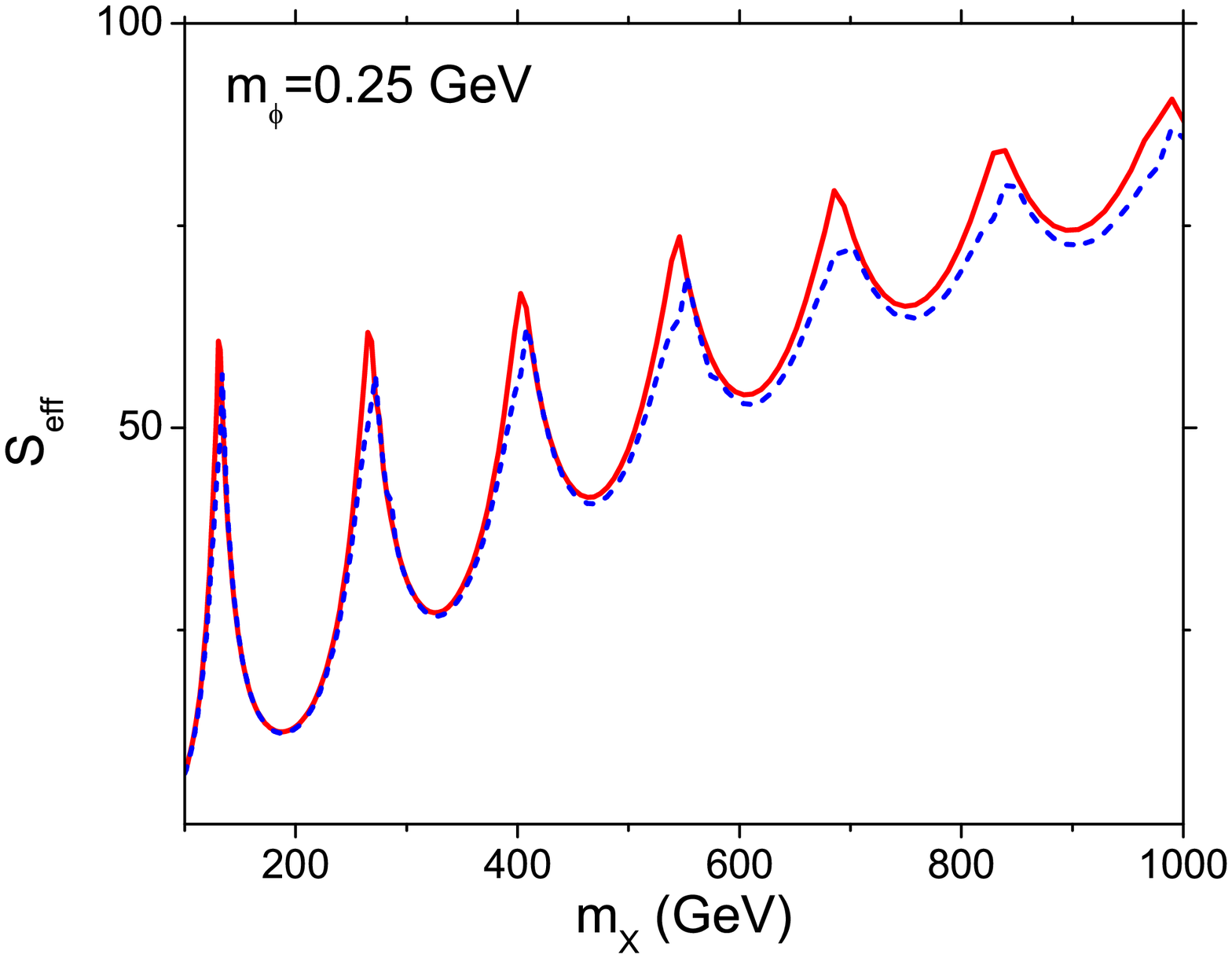}
\includegraphics*[width=0.49\columnwidth]{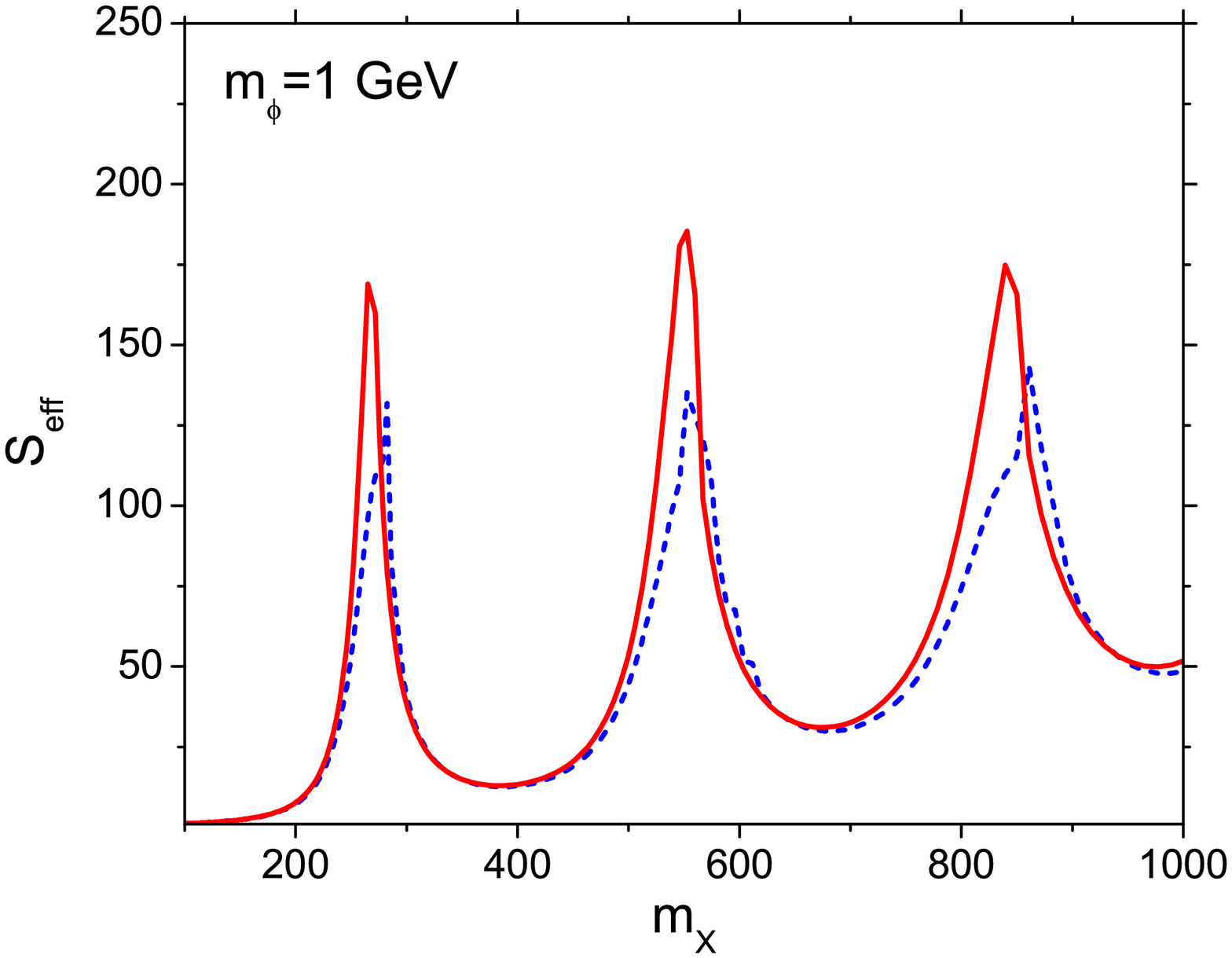}
\end{center}
\vspace*{-.25in}
\caption{The effective Sommerfeld enhancement factor $\Seff$ for $\Tkd
= \Tkd^e$ (solid red) and $\Tkd = \Tkd^{\phi}$ (dashed blue), for
$\mphi = 250~\mev$ (left) and 1 GeV (right). All other
$\Seff$-maximizing assumptions have been made.  Near resonances,
$\Seff$ is sensitive to the temperature of kinetic decoupling.
\label{fig:Seff_Tkd}}
\end{figure}

What is the effect on $\Seff$ of deviating from the $\Seff$-maximizing
assumptions?  The effect of the choice of $\Tkd$ is highly sensitive
to how close one is to a resonance.  Far from a resonance, there is
little sensitivity to $\Tkd$ values between $\Tkd^{\phi}$ and
$\Tkd^e$.  As one approaches a resonance, however, $\Seff$ may vary by
$\sim 10\%$ or more. In particular, we found regions close to
resonances for $\mphi\sim \gev$ where $\Seff$ changed by about
30\%. This behavior is shown in \figref{Seff_Tkd}.  If one is not very
close to a resonance, the choice of $\xs$ has a small impact.  In
contrast, as discussed above, the assumption of $\Omega_X h^2 = 0.114$
has a large effect; for other choices, the maximal value of $\Seff$
scales as $(\Omega_X h^2)^{3/2}$.  Finally, as can be seen in
\figref{volcano}, different choices of $\alpha_X$ also decrease $\Seff
\propto \alpha_X^3$ by $\sim 20-30\%$.

For typical parameters that are not very close to a resonance, then,
we expect that assumptions different from those listed above would
produce a small decrease in $\Seff$ when away from resonances.  Note,
however, that very near a resonance, all of these choices become very
important, as we now discuss.

\subsection{Resonances and Chemical Recoupling}
\label{sec:recoupling}

Very near a resonance, the annihilation cross section becomes
sensitive to details of the dark matter's evolution to low velocities
at late times.  The effective Sommerfeld enhancement then becomes
highly sensitive to details of $\Tkd$, the velocity distribution of
dark matter after $\Tkd$, and the cutoff of resonant enhancements.  As
this is far from generic, we do not present details of these
dependences, but we note that, contrary to naive expectations, $\Seff$
is not maximized by sitting exactly at resonance.  In fact, exact
resonances lead to extremely efficient annihilation in the early
Universe and minimize current indirect signals.

\begin{figure}[tb]
\begin{center}
\includegraphics*[width=0.49\columnwidth]{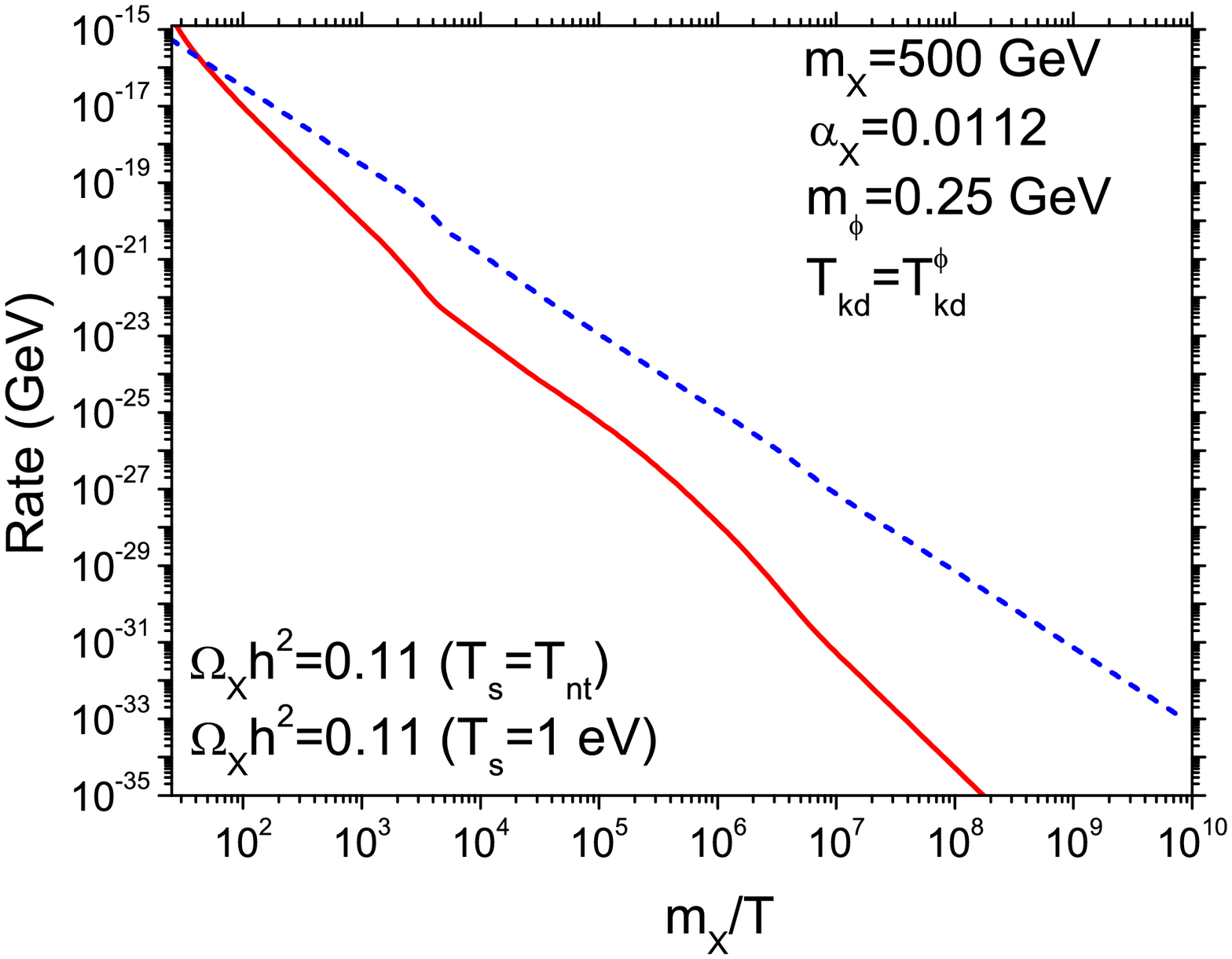}
\includegraphics*[width=0.49\columnwidth]{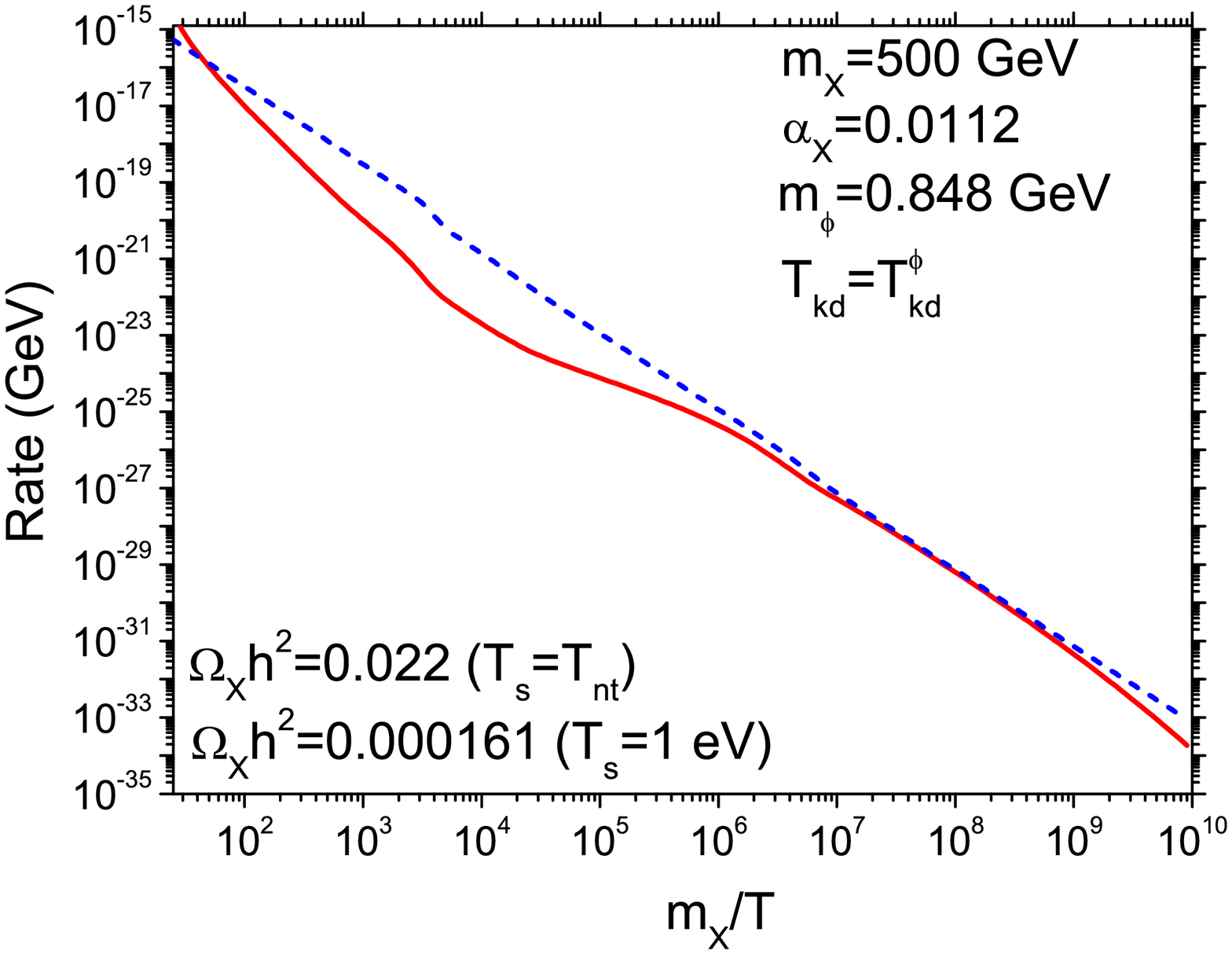}
\end{center}
\vspace*{-.25in}
\caption{The annihilation rate $\Gamma_{\text{an}} = n_X \langle
\sigmaan \vrel \rangle$ (solid red) and the Hubble rate $H$ (dashed
blue) as functions of $x = m_X/T$ for the values of $m_X$, $\mphi$,
$\alpha_X$ and $\Tkd$ indicated.  Without resonant Sommerfeld
enhancement (left), $\Gamma_{\text{an}} < H$ after freeze out; but
with resonant Sommerfeld enhancement (right), $\Gamma_{\text{an}}$ may
become comparable to $H$ at late times, leading to chemical recoupling
and an new era of annihilation. Also indicated on the plots are the
resulting values of $\Omega_Xh^2$ that result from assuming
annihilations up to a temperature of $\Tnt$ when the dark matter
distribution function is no longer able to maintain kinetic
equilibrium, as well as a lower temperature of 1 eV.
\label{fig:recoupling}}
\end{figure}

Resonant Sommerfeld enhancement's effect on freeze out can also be so
large that it leads to the intriguing phenomenon of chemical
recoupling.  This is illustrated for two representative cases in
\figref{recoupling}.  Without resonant Sommerfeld enhancement, dark
matter freezes out and remains frozen out, but with resonant
enhancement, dark matter may melt back in at late times, or chemically
recouple, leading to a second era of efficient annihilation.

The phenomenon of chemical recoupling in the context of Sommerfeld
enhancements may be understood as follows.  The Hubble parameter
scales as $H \sim T^2$.  The annihilation rate is $\Gamma_{\text{an}}
= n_X \langle \sigmaan \vrel \rangle$. The dark matter density scales
as $n_X \sim T^3$, and $\langle \sigmaan \vrel \rangle$ scales as
$\vrel^{-1} \sim T_X^{-1/2}$ off resonance and $\vrel^{-2} \sim
T_X^{-1}$ on resonance.  Before $\Tkd$, $T_X = T$, and so the
annihilation rate cannot grow relative to the expansion rate.
However, after $\Tkd$, $T_X \propto T^2$, and so $\Gamma_{\text{an}}
\sim T$.  At late times, then, the annihilation rate decrease slowly
enough to become comparable to or greater than the expansion rate. The
dark matter then melts back in, and there is a new era of
annihilation. In these cases, the relic density is very sensitive to
the temperature at which we stop including the annihilation
process. As our default case, we have been conservative and stopped
the annihilation process when the self-interactions of dark matter are
no longer able to maintain kinetic equilibrium. Annihilations will
proceed beyond this temperature; however, we cannot assume a
Maxwellian distribution for the dark matter particle, because the
annihilations preferentially deplete the low momentum tail. As an
example, we show in \figref{recoupling} what happens when we allow the
annihilation process to proceed down to 1 eV. The relic density is
essentially negligible for the case where we have chemical recoupling,
whereas it is unchanged for the case away from resonance.

\section{Comparison to PAMELA and Fermi}
\label{sec:comparison}

\subsection{Maximal Enhancements and Best Fit Parameters}

\begin{figure}[tbp]
\begin{center}
\includegraphics*[width=0.63\columnwidth]{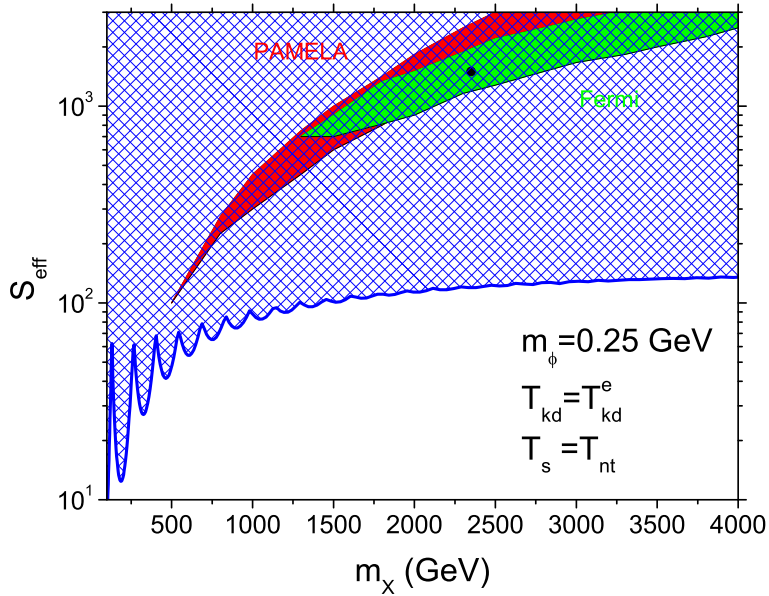}
\end{center}
\vspace*{-.25in}
\caption{ The maximal effective Sommerfeld enhancement factor $\Seff$
compared to data for force carrier mass $\mphi = 250~\mev$. The
cross-hatched region is excluded by requiring consistency with the
thermal relic density and adopting all of the $\Seff$-maximizing
assumptions listed in \secref{maximal}.  The red and green shaded
regions are $2\sigma$ PAMELA- and Fermi-favored regions for the $4\mu$
channel, and the best fit point is $(m_X, \Seff) = (2.35~\tev,
1500)$~\cite{Bergstrom:2009fa}.
\label{fig:fit}}
\end{figure}

In \figsref{fit}{fit1gev}, we compare the maximal $\Seff$ presented in
\figref{Seff_mx} to the boost factors required to explain PAMELA and
Fermi data.  The PAMELA and Fermi regions, derived in
Ref.~\cite{Bergstrom:2009fa}, assume an isothermal halo and a 250 MeV
force carrier that decays with 100\% branching ratio to $\mu^+ \mu^-$,
leading to annihilations $XX \to \mu^+ \mu^- \mu^+ \mu^-$.  The fits
are insensitive to $\mphi$, provided $\mphi \ll m_X$.  The best fit is
for $(m_X, \Seff) = (2.35~\tev, 1500)$, and the regions are $2\sigma$
contours relative to the best fit parameters.  Results for other halo
profiles, other particle physics models, and other final states have
also been
considered~\cite{Bergstrom:2009fa,Meade:2009iu,Abazajian:2010sq}.
These studies find that the 4$\mu$ final state provides a better fit
than all other considered final states, but see \secref{channels} for
a discussion of this issue.

\begin{figure}[tbp]
\begin{center}
\includegraphics*[width=0.63\columnwidth]{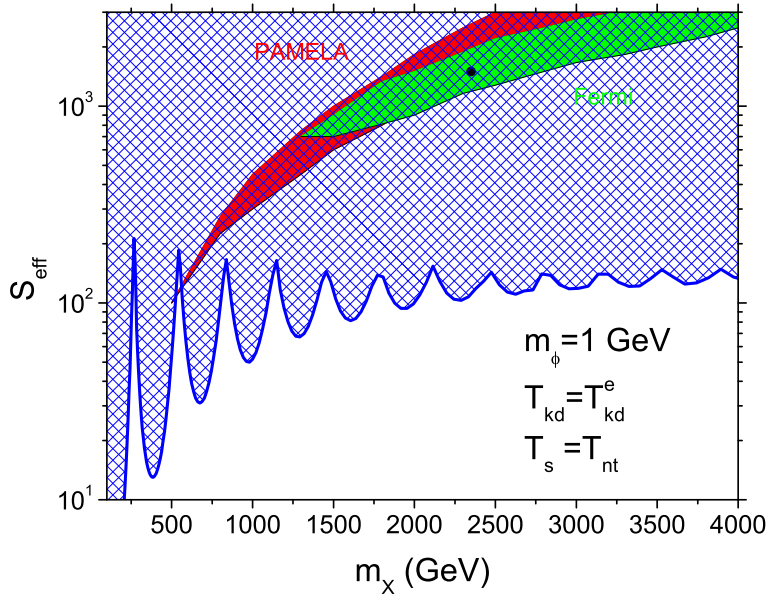}
\end{center}
\vspace*{-.25in}
\caption{The maximal effective Sommerfeld enhancement factor $\Seff$
compared to data for $\mphi=1~\gev$. The cross-hatched and shaded
regions are as in \figref{fit}. The PAMELA- and Fermi-favored regions
are for $\mphi=250~\mev$, but do not change appreciably for
$\mphi=1~\gev$~\cite{Meade:2009iu} in the case that only the $4\mu$
channel is included.
\label{fig:fit1gev}}
\end{figure}

Despite choosing $\Tkd$ and all other parameters to maximize $\Seff$,
we find that the possible values of $\Seff$ fall short of explaining
the PAMELA and Fermi excesses.  For example, at $m_X = 2.35~\tev$ the
maximal Sommerfeld enhancement is $\Seff \approx 100$, a factor of 15
below the best fit value.  For lower $m_X$, the data may be fit with
smaller $\Seff$, but the relic density bounds are also stronger; for
example, for $\mphi = 250~\mev$ and $m_X = 0.1$, 0.3, and 1 TeV, the
maximal values of $\Seff$ are 7, 30, and 90, respectively.

\subsection{Astrophysical Uncertainties}
\label{sec:astrophysics}

Our analysis so far has assumed standard astrophysics and
astroparticle physics.  To enable Sommerfeld enhancements to explain
PAMELA and Fermi, one may consider uncertainties in these assumptions.

The annihilation signal is sensitive to the square of the dark matter
density $\rho_X$ in the local neighborhood.  The best fit regions
given in Ref.~\cite{Bergstrom:2009fa} assume the conventional value
$\rho_X = 0.3~\gev/\cm^3$, which has traditionally been considered to
be uncertain up to a factor of 2~\cite{Kamionkowski:1997xg}.  More
recently, studies have tried to refine the calculation of the
uncertainty in the local density of dark matter. One study found
$\rho_X = 0.389 \pm 0.025~\gev/\cm^3$~\cite{Catena:2009mf}, using an
ensemble of kinematic tracers to constrain a standard galactic model
with a spherical halo, stellar and gas disk and stellar bulge. Another
recent study~\cite{Weber:2009pt} finds $\rho_X = 0.2 -
0.4~\gev/\cm^3$.  A third study Taylor-expanded the local rotation
velocity curve to first order and used local kinematic tracers and an
estimate of the baryonic contribution to the rotation curve to find
$\rho_X = 0.430 \pm 0.113 \pm 0.096~\gev/\cm^3$~\cite{Salucci:2010qr},
where the first uncertainty is from the slope of the circular velocity
$V(r)$ at the Sun's radius and the second is from uncertainty in
distance of the Sun from the galactic center. The three studies are
consistent with each other, but the first one finds significantly
smaller uncertainty in the value of $\rho_X$.  If we use the results
of Ref.~\cite{Catena:2009mf}, then the best fit regions plotted in
\figsref{fit}{fit1gev} shift down by a factor of $(0.4/0.3)^2$.  Of
course, it is not appropriate to simply scale up the signal by some
$\rho_X^2$ required to fit the positron and electron data; in a proper
analysis, one should include the appropriate likelihood for $\rho_X$
and $\Omega_X h^2$.  We have also discussed some of the uncertainties
in the velocity distribution function and their effects in
\Secref{maximal}, and, as with $\rho_X$ and $\Omega_Xh^2$, these
uncertainties should be included in the full likelihood analysis.

Another effect that was pointed out recently~\cite{Cline:2010ag} is
that the positrons produced within the unseen bound subhalos out in
the Milky Way dark matter halo could propagate to the local
neighborhood and contribute to the PAMELA and Fermi signal. For the
best fit point, this study found that about $30\%$ of the positrons in
PAMELA could be due to annihilations in the
subhalos~\cite{Cline:2010ag}. Taking $\rho_X = 0.4~\gev/\cm^3$ and an
enhancement factor of 30\% from subhalos, the discrepancy is reduced
from a factor of 15 to a factor of 4.  We see that these changes are
insufficient to reach the best fit regime, even if all the parameters
are simultaneously pushed in the most optimistic direction. A more
troubling aspect of this calculation is however the fact that the
subhalos considered were those resolved by the Via Lactea II
simulation. This implies that the much larger contribution from the
lower mass subhalos would over produce the signal. Appealing to
subhalos to ${\cal O}(1)$ changes to the required $\Seff$ is therefore
unmotivated.

Finally, we may also appeal to small scale structure in the immediate
local neighborhood to enhance the positron signals. Such a
contribution is largely unconstrained by data, but present
simulations, even without the deleterious effect of the disk of stars,
do not predict large enhancements from such an
effect~\cite{Vogelsberger:2008qb}. A recent study~\cite{Kuhlen:2009is}
using the Via Lactea II simulation and including the Sommerfeld effect
for annihilation in the local clumps found that the probability of
finding a subhalo close enough to make an ${\cal O}(1)$ effect on the
$e^+e^-$ flux above about 100 GeV is about 4\% and that this
contribution could be about 15\% if the Via Lactea II subhalo mass
function is extrapolated down to lower (unresolved) masses.  Two
points are worthy of further note here. First, these subhalo results
are going to be crucially affected by the disk. The zeroth order
expectation is that the gravity of the stellar disk will not allow
such subhalos to survive or even if they do, truncate it
significantly. Second, for small $\mphi$ and near resonances
self-interactions are important~\cite{Feng:2009hw,Buckley:2009in} and
that could modify the internal structure of the subhalo as well as the
mass function.

A more complicated set of astrophysical uncertainties is related to
astrophysical backgrounds and the signals themselves. For example, the
required Sommerfeld enhancement may be reduced if conventional
astrophysical backgrounds have been under-estimated, cosmic ray
propagation models are modified, pulsars contribute significantly to
the positron flux, or the PAMELA $e^+$ sample includes some proton
contamination.  Any of these effects could reduce the required dark
matter signal, but they could also plausibly eliminate the need for
dark matter altogether, as has been argued in numerous
studies~\cite{Hooper:2008kg,Yuksel:2008rf,Profumo:2008ms,%
Dado:2009ux,Biermann:2009qi,Katz:2009yd}.  It is therefore typically
unclear what the motivation would be for assuming that such effects
account for part of the signal but not all of it, independent of a
desire to leave room for a dark matter signal.

A possible exception to the above argument is the possibility of
adjusting cosmic ray propagation models to match the Fermi
spectrum. Such adjustments could remove the need to explain Fermi,
without changing the PAMELA excess
significantly~\cite{Grasso:2009ma}. As shown in \figref{fit1gev}, for
$m_X \alt~\tev$ and $\mphi \agt 1~\gev$, the allowed Sommerfeld
enhancements may be in marginal agreement with PAMELA data. This
scenario is, however, problematic. A contribution to the $e^\pm$ flux
at about 100 GeV to explain the observed PAMELA positron fraction
implies a modification to the $e^\pm$ flux observed by Fermi. For
example, a 500 GeV dark matter explanation for PAMELA with the correct
relic density is marginally possible for $\mphi=1~\gev$ according to
the $4\mu$ fits of Ref. \cite{Meade:2009iu} as shown in
\figref{fit1gev}.  This would, however, imply a drop in the $e^\pm$
flux beyond 500 GeV, which is not seen by Fermi, a point stressed in
previous works~\cite{Meade:2009iu,Abazajian:2010sq}. Analysis of such
a scenario with modified propagation models would be another avenue
for further study.

\subsection{Cosmic Microwave Background Bounds}
\label{sec:cmbbound}

In this section, we include bounds from the effect of residual
annihilation at last scattering on the ionized electron fraction and
therefore on the cosmic microwave background (CMB) power spectrum.
The CMB bound may be written as~\cite{Galli:2009zc,Slatyer:2009yq}
\begin{equation}
S|_{v=0} = \frac{12}{\epsilon_\phi
\left[ 1 - \cos\left( \sqrt{24 / \epsilon_\phi} \, \right)\right]}
< \frac{120}{f}\left(\frac{m_X}{\tev}\right) \ ,
\label{eq:cmbbound}
\end{equation}
where $f$ is the average fraction of the energy produced in
annihilation that reionizes Hydrogen between redshifts of 800 and
1000. For the $e^\pm$ final state, $f\simeq 0.7$, while for the
$\mu^\pm$ final state, $f\simeq 0.25$~\cite{Slatyer:2009yq}. The left
hand side $S|_{v=0}$ is the saturated Sommerfeld enhancement obtained
by setting $v=0$ in \eqref{Sapprox}. This is sufficient to approximate
the Sommerfeld enhancement during recombination, unless one is right
on top of a resonance, which is highly disfavored by this bound.

\Eqref{eq:cmbbound} is not a monotonic function of $\alpha_X$. To
impose this bound, we consider a region in $\alpha_X$ that is bounded
by values 10\% larger and smaller than the value of $\alpha_X$
dictated by the relic density bound of $\Omega_X h^2 = 0.114$.  We
find that large resonances are no longer allowed by the CMB bound as a
comparison of \figref{fit1gev} and \figref{constraints1gev} will show.

\subsection{Force Carrier Decay Channels}
\label{sec:channels}

The best fit regions of the $(m_X, \Seff)$ plane shown in
\figsref{fit}{fit1gev} assume that the force carrier decays solely to
muons, leading to the $X X \to \phi \phi \to 4\mu$ annihilation
channel.  The $4\mu$ final state has been
shown~\cite{Bergstrom:2009fa,Meade:2009iu,Barger:2009yt,Abazajian:2010sq}
to be the final state that yields the best fit of all considered so
far, as it leads to smooth $e^{\pm}$ distributions that can
simultaneously explain both PAMELA and Fermi.  At the same time, this
is an inefficient mode, as significant energy is lost to neutrinos.  To
ameliorate the discrepancy between the required $\Seff$ to explain the
anomalies and the maximal $\Seff$ allowed by the relic density
constraint, one might consider other channels, which could potentially
allow similar contributions to the $e^{\pm}$ spectrum with lower
$\Seff$ and at lower $m_X$.

The existence of other decay channels is theoretically well-motivated
and there are many possibilities.  If the decay $\phi \to \mu^+ \mu^-$
is possible, the decay $\phi \to e^+ e^-$ is also kinematically
allowed.  For $\mphi \sim 1~\gev$, other decay modes, such as $\phi
\to \pi \pi, K \bar{K}$, are also possible.  As an example, if $\phi$
decays are controlled by its kinetic mixing with the standard model
photon, then there will also be $e^\pm$ and $\pi^\pm$ channels with
roughly similar branching ratios, depending on
$\mphi$~\cite{Falkowski:2010cm}.  Alternatively, $\phi$ particles may
decay preferentially to the heaviest available states if they are
coupled through standard model Yukawa couplings, or may have other
interesting dynamics.  In fact, in full generality, the $\phi$
particles may decay not only to pairs of standard model particles, but
also through final states involving hidden Higgs bosons and other
hidden particles, leading, for example, to $XX$ annihilations to 6 or
8 particle final states.

Studies so far have found that the $4\mu$ mode is a better fit than
other channels~\cite{Meade:2009iu,Barger:2009yt}.  A precise
quantitative statement requires inclusion of correlations through
covariance matrices for each experiment, and these are not available.
Given this fundamental uncertainty, conclusions about the $\mu$ mode
and other channels are far from firm and require additional
assumptions.

In Ref.~\cite{Barger:2009yt}, the $4e$ mode is shown to have a
$\chi^2/\text{dof}$ that is larger by about $\sim 1$ compared the
$4\mu$ mode. These values of $\chi^2$ are computed by assuming that
the data points are uncorrelated and allowing for the spatial
diffusion parameter to vary. Other results obtained by fixing the
diffusion parameter seem to agree qualitatively with this
result~\cite{Meade:2009iu}. In fact, marginalizing over $m_X$ and
other parameters, Ref.~\cite{Meade:2009iu} found that the possibility
of mixed modes in the ratio of $4e:4\mu = 1:1$ is excluded at 99.9\%
CL.

\begin{figure}[tbp]
\begin{center}
\includegraphics*[width=0.63\columnwidth]{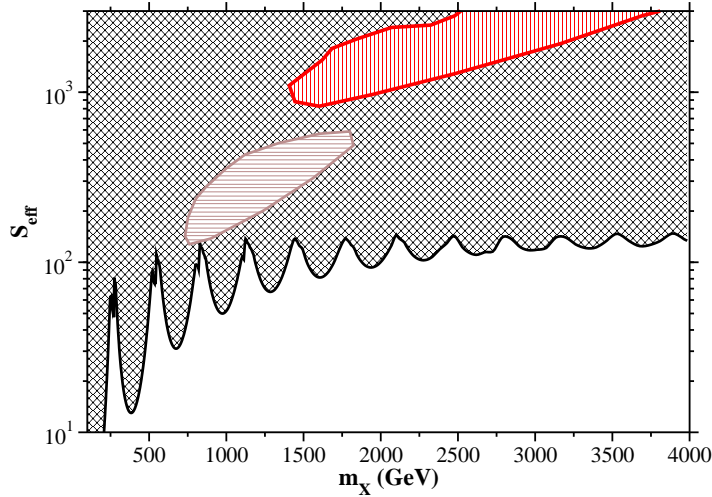}
\end{center}
\vspace*{-.25in}
\caption{The maximal effective Sommerfeld enhancement $\Seff$ (black
solid curve) after including bounds from CMB (see \secref{cmbbound})
compared to fits for $\mphi=1~\gev$ for exclusive $4e$ and $4\mu$
modes. The dark red (vertically hatched) region is the 99\% contour
for the $4\mu$ mode and the light red (horizontally hatched) contour
is for the $4e$ mode~\cite{Meade:2009iu}. The best fit regions for the
$4e$ mode are at lower $\Seff$ and $\m_X$, but the best fit $4e$ mode
point has a significantly lower likelihood than the best fit $4\mu$
mode point~\cite{Meade:2009iu} for a fixed cosmic ray propagation
model. \label{fig:constraints1gev}} 
\end{figure}

Despite these results, we may consider what impact other modes might
have, especially in light of the fact that there is a lot of freedom
in the cosmic ray propagation model parameters.  As an illustration,
we consider the best fit regions from Ref.~\cite{Meade:2009iu} for
both the $4e$ and $4\mu$ modes separately. We plot these best fit
regions with the relic density and CMB upper limits on $\Seff$ in
\figref{constraints1gev}.  The best-fit point for the $4e$ mode is at
lower $\Seff$ and $m_X$ and the edge of the region is in marginal
agreement with the relic density constraint.  Of course, as discussed
above, the best-fit point for the $4e$ mode has a much lower
likelihood than the best-fit point for the $4\mu$ mode, according to
Ref.~\cite{Meade:2009iu}.  For a 1:1 ratio, the best fit point would
move to $X$ masses somewhere between 1.5 and 2 TeV. This is disfavored
by the relic density requirement even if we set the local dark matter
density to be $0.4~\gev/\cm^3$.

For the reasons noted above, all of these conclusions are subject to
many uncertainties. The present conclusion in the literature seems to
that significant deviations from the pure $4\mu$ channel do not
provide a good fit to data. One way forward would be to redo fits of
the kind discussed in literature, allowing for variations in cosmic
ray propagation model parameters to see if this conclusion holds up in
general.  In any case, a careful statistical analysis is required
before appeals to alternative channels alone may be considered as a
way to allow Sommerfeld scenarios to explain the data.

\subsection{Bound States}

For the case of attractive forces, pairs of dark matter particles may
form bound states by radiating force carrier particles if the force
carrier mass $\mphi$ is below the binding energy $\alpha_X^2 m_X /
4$~\cite{Pospelov:2008jd,MarchRussell:2008tu,Shepherd:2009sa}.  As in
the case of positronium, when this is kinematically possible,
annihilation through bound states is preferred.  In the limits $\mphi
\ll \alpha_X^2m_X/4$ and $v \gg \mphi/m_X$, the annihilation rate
enhancement is about $20 \alpha_X v_\phi(3-v_\phi^2)/(2v)$ for
fermions, where $v_\phi$ is the velocity of the final state $\phi$
particle when the bound state forms~\cite{Pospelov:2008jd}. This is
roughly a factor of 7 larger than $S^0$ when $v_\phi \simeq 1$.

Unfortunately, such bound state effects do not help Sommerfeld
enhancement models explain the PAMELA and Fermi data for two reasons.
First, bound state formation is kinematically forbidden in large parts
of parameter space that yield the best fits to data, and the large
factor of 7 enhancement alluded to above is not realized for most of
the parameter space plotted in \figref{fit1gev}. To see this, we first
need to compute the value of $\alpha_X$ required to get the right
relic density. Now, bound states may form in the early Universe when
they are energetically viable, that is when $v < \alpha_X$, and hence
affect the relic density calculation. For simplicity we neglect this
effect here, and use the same $\alpha_X$ values from \figref{alphamx},
which leads to a conservative bound.  Then, for $\mphi = 0.25~\gev \
(1~\gev)$, the bound state enhancement formula quoted above is valid
only for $m_X \gg 1.5~\tev \ (2.5~\tev)$. These inequalities are not
satisfied for parameters that are most promising for fitting the data
($\mphi \agt 0.25~\gev$ and $m_X \alt 4~\tev$).

Second, if bound states form, they are likely to be in conflict with
bounds from the CMB. To see this, let's saturate the bound state
enhancement at $v\sim \mphi/m_X$ and estimate the enhancement at
recombination as $20\alpha_X m_X/\mphi$. This should be compared to
the CMB bound of $(120/f) (m_X/\tev) \simeq 500~(m_X / \tev)$, where we
have set $f=0.25$ to be conservative. To
satisfy the CMB bound, then, we require $\alpha_X < (\mphi/\gev) /
40$.  This is generically violated if $\alpha_X$ is fixed by the relic
density. For example, for $\mphi=1~\gev$, the bound implies $\alpha_X
< 0.025$.  Requiring the correct relic density then implies $m_X <
1~\tev$. However, we need $m_X > 2.5~\tev$ for bound states to form
with $\mphi=1~\gev$. The situation is no different with $\mphi =
0.25~\gev$.  Thus our saturation approximation above (for bound state
enhancement) suggests that regions of parameter spaces where bound
states can form are generically in conflict with the CMB.

\subsection{Non-minimal Particle Physics Models}
\label{sec:particlephysics}

A potentially more promising direction is to construct more
complicated particle physics models to achieve Sommerfeld enhancements
that are large enough to explain PAMELA and Fermi.  Such models will
generically include additional annihilation channels.  We begin by
discussing the impact of these channels in general, and then examine
various strategies one might explore to achieve larger $\Seff$.

\subsubsection{Additional Annihilation Channels}

As discussed above, to maximize $\Seff$, we have considered only $XX
\to \phi \phi$ annihilation, leading to the typical tree-level cross
section estimate of \eqref{sigma}.  Even in the simplest models,
however, one may have additional annihilation channels.  For example,
if $\phi$ is a U(1) gauge boson with mass generated by spontaneous
symmetry breaking through the Higgs potential $\frac{1}{2} \lambda (
|H|^2 - v^2 ) ^2$, where $H$ is a complex scalar, there is a physical
hidden Higgs boson $h$ with mass $m_h \sim \sqrt{2 \lambda} \, v$,
where $v$ is related to the $\phi$ mass by $\mphi = \sqrt{8 \pi
\alpha_X} \, v$.  For perturbative $\lambda$, $m_h \alt 10 \, \mphi$;
$h$ may be much lighter than $\phi$, but it cannot be much heavier.
The hidden Higgs boson therefore cannot be decoupled either
dynamically or kinematically, and there is necessarily an additional
annihilation channel $XX \to \phi h$.  In the non-relativistic limit,
and assuming there is no Yukawa coupling $h \bar{X} X$, the cross
section is
\begin{equation}
\sigma(X X \to \phi h) \vrel =
\frac{\pi \alpha_X^2}{m_X^2} \frac{|\bar{p}_{\phi}|}{m_X}
\frac{4 m_X^4 + 8 m_X^2 \mphi^2}{(4 m_X^2 - \mphi^2)^2} \ ,
\end{equation}
where
\begin{equation}
|\bar{p}_{\phi}| =
\frac{\left[ \left( 4 m_X^2 - (\mphi + m_h)^2 \right)
\left( 4 m_X^2 - (\mphi - m_h)^2 \right) \right]^{1/2} } {4 m_X} \ .
\end{equation}
In the limit $\mphi, m_h \ll m_X$,
\begin{equation}
\sigma(X X \to \phi h) \vrel =
\frac{1}{4} \frac{\pi \alpha_X^2}{m_X^2} \ .
\end{equation}
Additional annihilation channels and their impact on relic densities
in Sommerfeld enhanced models have also been considered in, for
example, Ref.~\cite{MarchRussell:2008yu}.

\begin{figure}[tb]
\begin{center}
\includegraphics*[width=0.49\columnwidth]{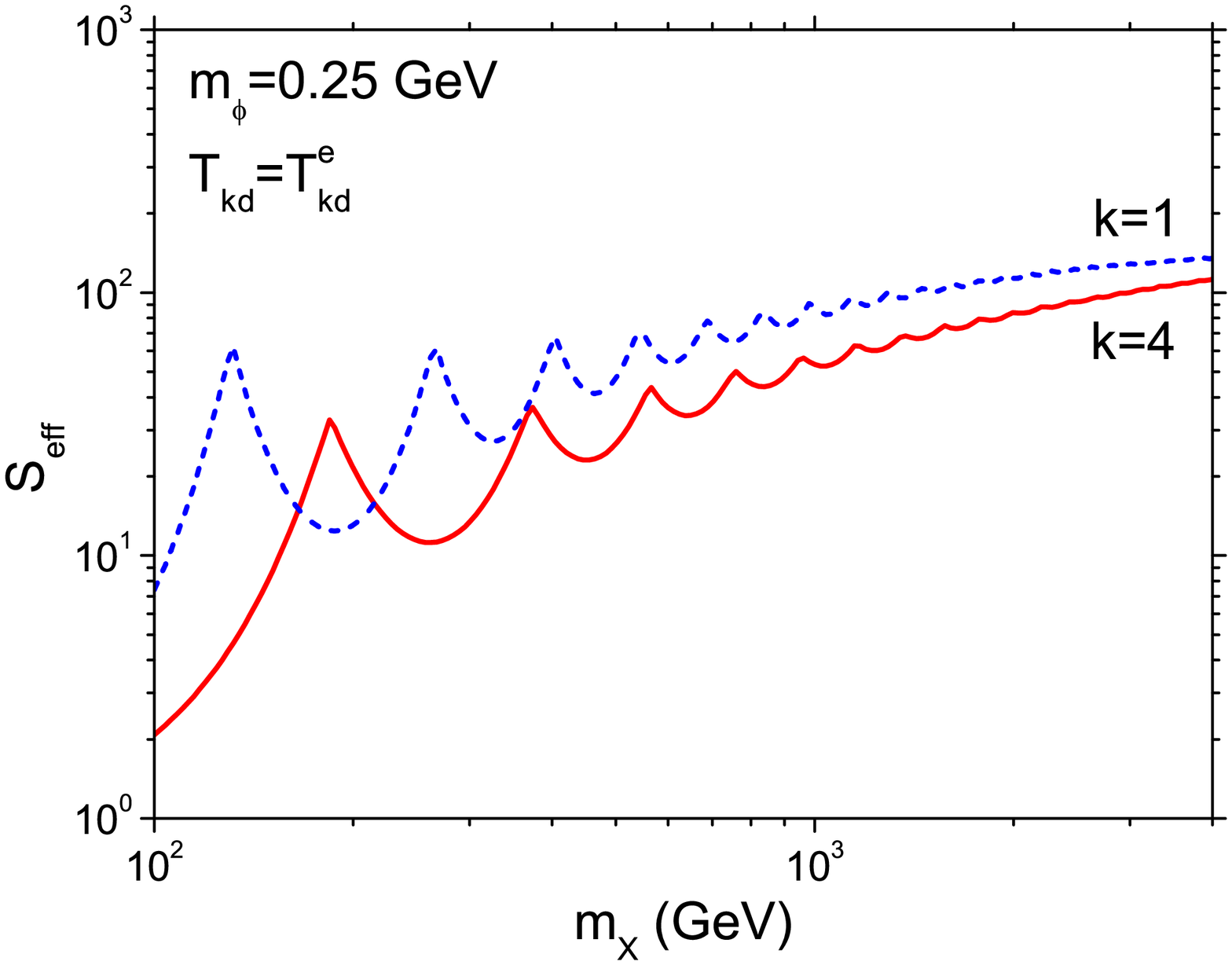}
\includegraphics*[width=0.49\columnwidth]{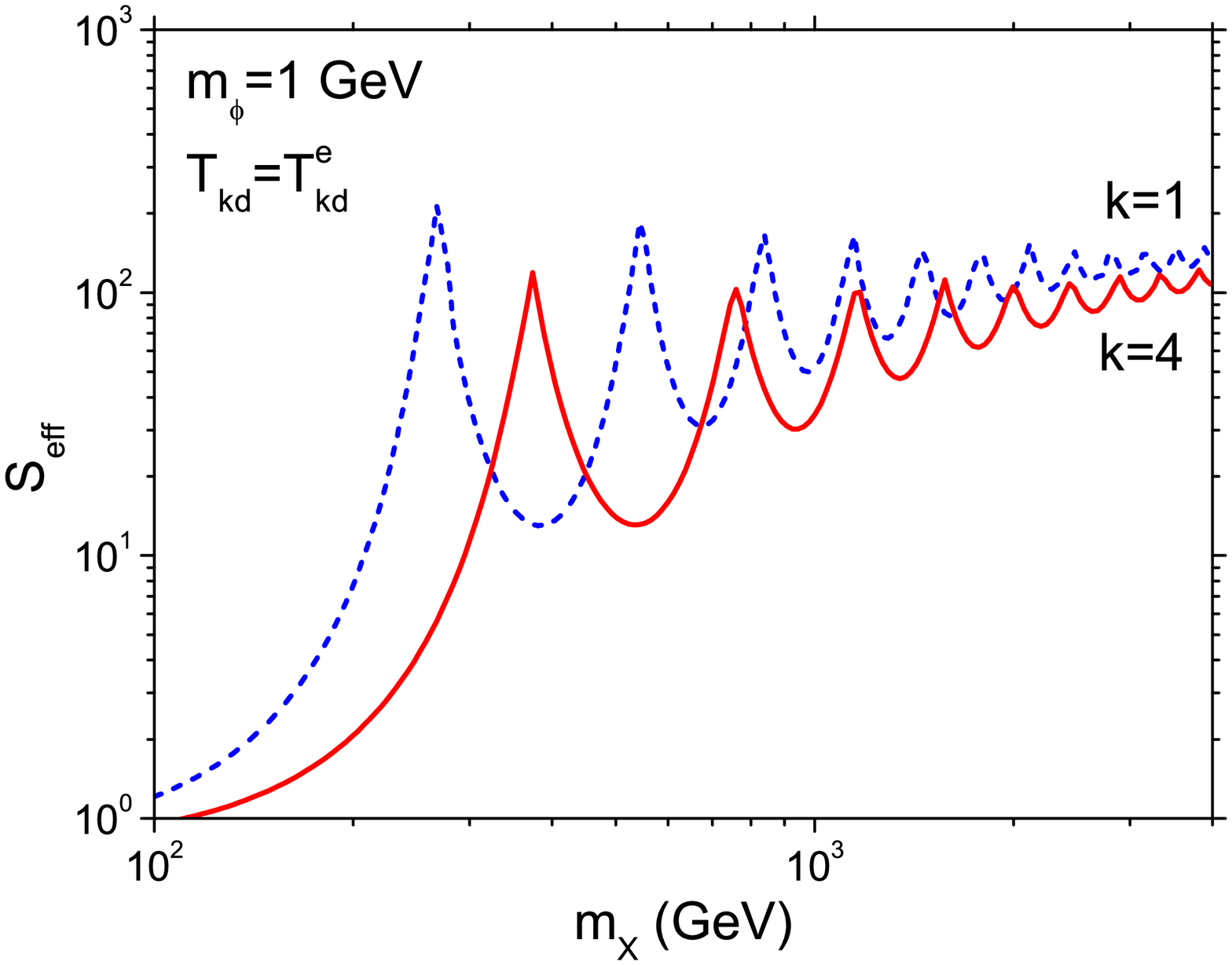}
\end{center}
\vspace*{-.25in}
\caption{The maximal effective Sommerfeld enhancement factor $\Seff$
for $k = 1, 4$, where the tree-level annihilation cross section is
$\sv=k\pi\alpha_X^2/m_X^2$, for $\mphi=0.25~\gev$ (left) and
$\mphi=1~\gev$ (right). The effects scale roughly as $1/\sqrt{k}$,
with the deviation from $1/\sqrt{k}$ scaling due to the Sommerfeld
effect in the early Universe.
\label{fig:kfigs}}
\end{figure}

To roughly accommodate such modifications, we may parameterize the
effect of additional annihilation channels by assuming a tree-level
annihilation cross section
\begin{equation}
\sv = k \frac{\pi \alpha_X^2}{m_X^2} \ ,
\end{equation}
where $k$ is a constant.  If the new structures do not significantly
modify the Sommerfeld enhancement $S$, the desired relic density will
be achieved by modifying $\alpha_X \to \alpha_X / \sqrt{k}$, moving
the positions of resonances and reducing the maximal $\Seff$ by the
factor $\sqrt{k}$. In general, Sommerfeld enhancement in the early
Universe is important and the reduction in maximal $\Seff$ is smaller.
The effects of varying $k$ are shown in \figref{kfigs}.

Of course, in general, new annihilation channels may have different
couplings $\alpha_X$, different kinetic decoupling features, and
different $\Tnt$, all leading to different Sommerfeld effects,
resonances in different places, and many other effects that cannot be
captured by a single extra parameter $k$.  The availability of
additional annihilation channels also has a complicated effect on the
signal.  For example, for the case of decays to Higgs bosons and
assuming $h \to 4l$ through real or virtual $\phi$ pairs, the $h \phi$
mode leads to six or more leptons.  This channel changes the energy
spectrum of positrons that would be seen by PAMELA and Fermi and thus
requires new fits to the data.  The full analysis is therefore
complicated, and requires a dedicated study incorporating all of these
effects into the freeze out analysis.  The main point is that
non-minimal models will typically include additional hidden sector
states and additional annihilation channels, generically reducing the
maximum possible $\Seff$ consistent with thermal relic density
constraints.

\subsubsection{Multi-state Dark Matter}

Two-state dark matter models have received considerably attention
recently.  For example, ``inelastic dark matter'' has been motivated
by the DAMA NaI signal~\cite{TuckerSmith:2001hy,Chang:2008gd} and
``exciting dark matter'' has been proposed to explain the INTEGRAL/SPI
excess towards the galactic center~\cite{Finkbeiner:2007kk}. They have
also been proposed as explanations of the PAMELA excess and the WMAP
haze towards the galactic
center~\cite{Finkbeiner:2009mi,Cholis:2008vb}.  In these models, the
stable dark matter state is accompanied by a more massive unstable
state, which is separated from the stable state by a mass gap that is
chosen to fit the data and is much smaller than the dark matter
mass. The presence of this almost degenerate excited state
significantly complicates the early Universe freeze out analysis. The
Sommerfeld enhancements are also different~\cite{Slatyer:2009vg}. To
compute the relic density, one would need to include the additional
annihilation channels and the two-state dark matter Sommerfeld
enhancement in a self-consistent early Universe freeze out
calculation.

Dark matter models with multiple {\em stable} dark matter states have
also been proposed to explain the PAMELA and Fermi excesses and the
DAMA/LIBRA signal~\cite{Zurek:2008qg,Katz:2009qq,Cholis:2009va}. These
models provide an interesting avenue to reduce the required Sommerfeld
enhancement to explain both the PAMELA and Fermi
excesses~\cite{Cholis:2009va}. This is because the heavier ($\sim
\tev$) component contributes mainly to the Fermi excess, whereas the
lighter ($\sim 100~\gev$) dark matter particle contributes
predominantly to the PAMELA positron excess. The required Sommerfeld
enhancements are about 100 for the TeV state.

Such enhancements may appear to be consistent with our relic density
constraint to within a factor of 2, but the relic density calculation
that we have presented here is not complete in these scenarios.  In
the simplest such models~\cite{Cholis:2009va}, the two states couple
to the dark gauge boson with the same charge. As stated, however,
these models are ruled out by the direct detection limits, because the
$Xq\to Xq$ cross section due to the kinetic mixing would be too large
for $\epsilon >
10^{-6}$~\cite{Pospelov:2007mp,ArkaniHamed:2008qn}. The way out of
this constraint is to specify that each state is further split into a
metastable and stable state with very small mass gap and that the
$\phi$ particles only have off-diagonal couplings, so that kinetic
energy larger than the mass gap is required to initiate $X q \to
X^*q$, where $X^*$ denotes the excited state of
$X$~\cite{ArkaniHamed:2008qn}. These mass gaps may then be chosen to
be $\sim{\cal O}(\mev)$ or $\sim{\cal O}(0.1~\mev)$ to explain
INTEGRAL/SPI and DAMA/LIBRA signals respectively~\cite{Cholis:2009va}.
The relic density calculation is now significantly more
complicated. Additional annihilation channels, for example,
annihilations between the heavy and light states and annihilation to
$h\phi$, and the presence of an almost degenerate state must be taken
into account in the early universe calculation. The Sommerfeld effect
for the lighter state is generically important because the coupling to
the dark gauge boson would be set by the relic density constraint for
the heavy particle, and so $\alpha_X \sim 0.02$, much larger than the
required $\alpha_X$ for thermal dark matter with mass $\sim
100~\gev$. The new channels, such as the $h\phi$ channel, also modify
comparisons to PAMELA and Fermi as these new channels may contribute
to the $e^{\pm}$ flux with a different energy spectrum. These issues
are beyond the scope of the present work.

Multi-state dark matter theories may also have significantly
suppressed $k$.  For example, in the non-Abelian models of
Ref.~\cite{Chen:2009ab}, values of $k$ as small as $0.14$ are
possible, and these have recently been claimed to provide marginally
consistent explanations of the cosmic ray
anomalies~\cite{Cirelli:2010nh}.  In these studies, however, the
irreducible annihilations to Higgs bosons have not been included, and
the Sommerfeld effect on freeze out has been neglected.  Given the low
value of $k$, these changes to the annihilation cross section may be
important. It would be interesting to see if the consistency remains
after a complete analysis.

To summarize, finding a model with more degrees of freedom that
requires smaller boosts to fit the PAMELA and Fermi data does not by
itself imply an improvement, as the bounds on $\Seff$ will generically
also be stronger.  Generalizations of the freeze out analysis
described here are required to determine if these scenarios may
self-consistently explain PAMELA and Fermi data.

\subsubsection{Very Heavy Force Carriers with $\mphi \gg 1~\gev$}

One reason the maximal $\Seff$ is somewhat limited is that the
resonances are not significant for $\mphi \alt 1~\gev$ and $\m_X \agt
1~\tev$.  For larger $\mphi$, however, the resonances move to larger
$m_X$, as evident from \eqref{mphiresonance}.  Generically, for $\mphi
> 2~\gev$, $\phi$ decay may produce anti-protons.  However, if such
contributions to the anti-proton spectrum may be otherwise suppressed
or accommodated~\cite{Kane:2009if}, such large $\mphi$ may allow one
to improve the consistency of Sommerfeld-based explanations of the
PAMELA and Fermi data.

As an example, in Ref.~\cite{Fox:2008kb}, a leptophilic model was
constructed where the anti-proton flux is suppressed by dynamics,
freeing the force carrier mass to be much larger than $\sim \gev$.
For $\mphi \sim 10~\gev$ and $m_X \approx 800~\gev$, the authors noted
that resonant Sommerfeld enhancement could enhance signals with
enhancement factors as large as $S \sim 1000$.  However, the study of
Ref.~\cite{Fox:2008kb} did not include the Higgs annihilation channel,
and did not include the impact of Sommerfeld enhancement on freeze
out.  Note also that, as discussed in \secref{phidecay}, for very
large $\mphi$ near the freeze out temperature, $\phi$ particles may
not be in complete thermal equilibrium with the standard model,
modifying the freeze out analysis.  Nevertheless, such possibilities
certainly merit further study.  Note, though, that these models are
inaccessible to searches for GeV forces.

\subsubsection{Running $\alpha_X$}

In all of our analysis above, we have assumed that the running of the
gauge coupling in the dark sector is not significant. Although this is
certainly a good approximation for the minimal Abelian model we have
considered, the situation is much more complicated for non-Abelian
models. It has been recently pointed out that a non-Abelian hidden
sector could explain both PAMELA and Fermi, while resulting in the
correct relic density~\cite{Zhang:2009dd}. This work notes that the
Sommerfeld effect is a soft process and the enhancement factor $S$ is
determined by the coupling at scale $\mphi$, whereas for the tree-level
cross section the coupling must be evaluated at $m_X$. For a
non-Abelian theory where the coupling constant is smaller at higher
energies, this could significantly relax the tension between the relic
density constraint and the requirements for explaining PAMELA and
Fermi.

However, this analysis~\cite{Zhang:2009dd} did not include the
Sommerfeld effect on the annihilation in the early Universe. This
would be particularly important because the Sommerfeld enhancement is
now much larger even in the early Universe. The model also includes a
Higgs doublet $H$ and the dark matter annihilation $XX \to HZ$ must
also be included. The stability of the Higgs sector is a major concern
for these models because the Higgs annihilation cross section is too
small to suppress their relic density~\cite{Zhang:2009dd}. The decay
lifetimes that have been computed~\cite{Zhang:2009dd} are long ($\sim
10^6~\s$ or longer) and could destroy the successful predictions of
Big Bang nucleosynthesis (BBN) because of the injection of
electromagnetic energy~\cite{Kawasaki:1994sc,Cyburt:2002uv}. Given
these issues, it is not clear that these models can successfully
satisfy the relic density constraints, but future work addressing
these points could be an interesting way forward to explain PAMELA and
Fermi excesses using Sommerfeld-enhanced annihilation.

We note that the arguments regarding the decay of the hidden sector
Higgs boson are quite general. The hidden Higgs boson should decay to
standard model particles, otherwise it would contribute to the dark
matter abundance. In doing so, the decays should not destroy the
successful predictions of BBN.

\section{Conclusions}
\label{sec:conclusions}

In this work, we have investigated scenarios with Sommerfeld-enhanced
annihilation of thermal dark matter and determined the largest
possible indirect signals, subject to the most basic of constraints,
namely, that the thermal dark matter have the correct thermal relic
density.  In addition to their potential impact on cosmic $e^{\pm}$
spectra, such scenarios may also predict observable effects in the
cosmic microwave background and the diffuse extragalactic $\gamma$-ray
background, excess $\gamma$-ray, neutrino and radio wave emission from
the galactic center, and excess $\gamma$-rays from dwarf spheroidal
satellites of the Milky Way and extended regions centered on the
galactic center~\cite{Kamionkowski:2008gj,Belikov:2009qx,%
Galli:2009zc,Profumo:2009uf,Slatyer:2009yq,Huetsi:2009ex,%
Meade:2009iu,Cirelli:2009bb,Kanzaki:2009hf,Crocker:2010gy,%
Abazajian:2010sq,Calvez:2010jq,PalomaresRuiz:2010pn}. The absence of
each of these potential signatures provides significant
constraints. The bounds derived here are complementary to those, in
that these bounds are derived by studying not the annihilation
products, but the production.  These results are therefore
model-independent and robust, in that they assume only that the dark
matter is produced by thermal freeze out, which is the central
assumption motivating the consideration of Sommerfeld enhancements in
the first place.

This analysis included the possibility of resonant Sommerfeld
enhancement and the effect of Sommerfeld enhancement in the thermal
relic density calculation.  The possibility of Sommerfeld enhancement
at low velocities, with annihilation cross sections proportional to
$\vrel^{-1}$, or even $\vrel^{-2}$ on resonance, implies that the
thermal relic density is sensitive to the dark matter density's
evolution long after conventional freeze out at $T_f \sim m_X / 25$.
As a result, the relic density depends on many aspects absent in
conventional scenarios.  These include the cutoff of resonant
Sommerfeld annihilation, the equilibration of force
carrier particles, the temperature of kinetic decoupling $\Tkd$, and
the efficiency of self-scattering to thermalize the dark matter
velocity distribution after $\Tkd$.

These dependences may be very important.  A non-generic but intriguing
example is that, on resonance, dark matter freezes out but may then
later melt in, chemically recoupling for a second era of efficient
annihilation.  As a result, exact resonances suppress rather than
enhance indirect signals, in contrast to naive expectations.  More
generally, the Sommerfeld enhancement of freeze out implies that
smaller tree-level annihilation cross sections are required to give
the correct relic density, and so reduces the effective Sommerfeld
enhancement of indirect signals.  We have found reduction factors of a
few, with several $\sim 10 - 30\%$ variations possible, depending on
the value of $\Tkd$ and other freeze out parameters.

For the minimal scenario analyzed here, and adopting the most
optimistic assumptions to maximize indirect signals, the largest
possible values of $\Seff$ are 7, 30, and 90 for $\mphi = 250~\mev$
and $m_X = 0.1$, 0.3 and 1 TeV, respectively.  As shown in
\figsref{fit}{fit1gev}, such enhancements fall short of explaining the
PAMELA and Fermi cosmic $e^{\pm}$ excesses.  For the best fit point,
as discussed in \secref{comparison}, this discrepancy is over an order
of magnitude and cannot be eliminated by appeals to enhanced local
dark matter density or boosts from small scale structure.  Bound state
effects are also unable to eliminate the inconsistency, as they are
typically insignificant in the parameter regions that may potentially
explain the cosmic ray anomalies, and are in any case stringently
constrained by the CMB.  Along with the astrophysical uncertainties
discussed in \secref{comparison} and the complementary constraints
from the cosmic microwave background and other sources, these results
imply that the dark matter motivations for Sommerfeld enhancements and
GeV dark forces at present are, at best, strained.

We have discussed several possibly interesting directions for further
study in \secref{comparison}. Non-minimal models will generically
include additional annihilation channels, such as those involving
hidden Higgs bosons.  These will generically strengthen the bounds on
$\Seff$.  We noted also that the decay of the hidden sector Higgs
boson to standard model particles could be constrained by demanding
consistency with the measured light element abundances produced during
BBN.  Additional features and constraints particular to specific
models have been noted in \secref{comparison}.  We have focused on the
$4\mu$ final state here, which generically give the best fits to the
data. Different annihilation channels, such as $4e$ final states, are
optimized at lower $\Seff$ and $m_X$, but also yield worse fits to the
data according to present estimates in the literature. However, this
conclusion may depend significantly on the cosmic ray propagation
model parameters used. At present, no Sommerfeld models have been
shown to explain the PAMELA and Fermi excesses consistent with the
constraint of thermal freeze out, and we stress that complete and
dedicated studies of freeze out are required to judge properly whether
more complicated models can self-consistently explain PAMELA and Fermi
data.

The maximal Sommerfeld enhancements derived here, although not
sufficient to explain current PAMELA and Fermi excesses, are
nevertheless significant enhancements.  These will be probed as PAMELA
and Fermi continue to gather data, and, in the near future, at AMS on
the International Space Station.  The interpretation of future data as
dark matter may, of course, continue to be clouded by confusion from
astrophysical uncertainties, but the diverse set of particle physics
and astroparticle physics experiments that may probe weak-scale dark
matter in the near future at least provides some reason for optimism.

\section*{Acknowledgments}

We thank Masahiro Ibe, Jason Kumar, Erich Poppitz, and Neil Weiner for
useful conversations and correspondence, and Tracy Slatyer for
motivating more detailed investigation of resonant Sommerfeld
enhancements and comments on an early draft of this work. We thank
Marc Kamionkowski for pointing out the importance of inverse decays
for the force carrier in equilibrium, and the participants of the
Aspen Summer 2010 workshop ``From Colliders to the Dark Sector'' for
stimulating discussions.  The work of JLF and HY was supported in part
by NSF grants PHY--0653656 and PHY--0709742. The work of MK was
supported in part by NSF grant PHY--0855462 and NASA grant
NNX09AD09G. 




\begin{thebibliography}{99}

\bibitem{Goldberg:1983nd}
  H.~Goldberg,
  Phys.\ Rev.\ Lett.\  {\bf 50}, 1419 (1983)
  [Erratum-ibid.\  {\bf 103}, 099905 (2009)].

\bibitem{Feng:2009hw}
  J.~L.~Feng, M.~Kaplinghat and H.~B.~Yu,
  Phys.\ Rev.\ Lett.\  {\bf 104}, 151301 (2010)
  [arXiv:0911.0422 [hep-ph]].

\bibitem{Barwick:1997ig}
  S.~W.~Barwick {\it et al.}  [HEAT Collaboration],
  Astrophys.\ J.\  {\bf 482}, L191 (1997)
  [arXiv:astro-ph/9703192].

\bibitem{Beatty:2004cy}
  J.~J.~Beatty {\it et al.},
  Phys.\ Rev.\ Lett.\  {\bf 93}, 241102 (2004)
  [arXiv:astro-ph/0412230].

\bibitem{Adriani:2008zr}
  O.~Adriani {\it et al.} [PAMELA Collaboration],
  Nature {\bf 458}, 607 (2009) [arXiv:0810.4995 [astro-ph]].

\bibitem{:2008zzr}
  J.~Chang {\it et al.} [ATIC Collaboration],
  Nature {\bf 456}, 362 (2008).

\bibitem{Abdo:2009zk}
  A.~A.~Abdo {\it et al.} [The Fermi LAT Collaboration],
  Phys.\ Rev.\ Lett.\ {\bf 102}, 181101 (2009)
  [arXiv:0905.0025 [astro-ph.HE]].

\bibitem{Strong:2009xj}
  A.~W.~Strong \etal,
  arXiv:0907.0559 [astro-ph.HE].

\bibitem{Sommerfeld:1931}
A.~Sommerfeld,
Annalen der Physik {\bf 403}, 257 (1931).

\bibitem{Baer:1998pg}
  H.~Baer, K.~m.~Cheung and J.~F.~Gunion,
  Phys.\ Rev.\  D {\bf 59}, 075002 (1999)
  [arXiv:hep-ph/9806361].

\bibitem{Hisano:2002fk}
  J.~Hisano, S.~Matsumoto and M.~M.~Nojiri,
  Phys.\ Rev.\  D {\bf 67} (2003) 075014
  [arXiv:hep-ph/0212022].

\bibitem{Hisano:2003ec}
  J.~Hisano, S.~Matsumoto and M.~M.~Nojiri,
  Phys.\ Rev.\ Lett.\  {\bf 92}, 031303 (2004)
  [arXiv:hep-ph/0307216].

\bibitem{Hisano:2004ds}
  J.~Hisano, S.~Matsumoto, M.~M.~Nojiri and O.~Saito,
  Phys.\ Rev.\  D {\bf 71}, 063528 (2005)
  [arXiv:hep-ph/0412403].

\bibitem{Hisano:2005ec}
  J.~Hisano, S.~Matsumoto, O.~Saito and M.~Senami,
  Phys.\ Rev.\  D {\bf 73}, 055004 (2006)
  [arXiv:hep-ph/0511118].

\bibitem{Cirelli:2007xd}
  M.~Cirelli, A.~Strumia and M.~Tamburini,
  Nucl.\ Phys.\  B {\bf 787}, 152 (2007)
  [arXiv:0706.4071 [hep-ph]].

\bibitem{Cirelli:2008pk}
  M.~Cirelli, M.~Kadastik, M.~Raidal and A.~Strumia,
  Nucl.\ Phys.\  B {\bf 813}, 1 (2009)
  [arXiv:0809.2409 [hep-ph]].

\bibitem{ArkaniHamed:2008qn}
  N.~Arkani-Hamed, D.~P.~Finkbeiner, T.~R.~Slatyer and N.~Weiner,
  Phys.\ Rev.\  D {\bf 79}, 015014 (2009)
  [arXiv:0810.0713 [hep-ph]].

\bibitem{Pospelov:2008jd}
  M.~Pospelov, A.~Ritz,
  Phys.\ Lett.\  {\bf B671}, 391-397 (2009)
  [arXiv:0810.1502 [hep-ph]].

\bibitem{Fox:2008kb}
  P.~J.~Fox and E.~Poppitz,
  Phys.\ Rev.\  D {\bf 79}, 083528 (2009)
  [arXiv:0811.0399 [hep-ph]].

\bibitem{Bergstrom:2009fa}
  L.~Bergstrom, J.~Edsjo and G.~Zaharijas,
  Phys.\ Rev.\ Lett.\  {\bf 103}, 031103 (2009)
  [arXiv:0905.0333 [astro-ph.HE]].

\bibitem{Meade:2009iu}
  P.~Meade, M.~Papucci, A.~Strumia and T.~Volansky,
  Nucl.\ Phys.\  B {\bf 831}, 178 (2010)
  [arXiv:0905.0480 [hep-ph]].

\bibitem{Iengo:2009ni}
 R.~Iengo,
 JHEP {\bf 0905}, 024 (2009)
 [arXiv:0902.0688 [hep-ph]].

\bibitem{Iengo:2009xf}
 R.~Iengo,
 arXiv:0903.0317 [hep-ph].

\bibitem{Cassel:2009wt}
  S.~Cassel,
  arXiv:0903.5307 [hep-ph].

\bibitem{Slatyer:2009vg}
  T.~R.~Slatyer,
  JCAP {\bf 1002}, 028 (2010)
  [arXiv:0910.5713 [hep-ph]].

\bibitem{Hisano:2006nn}
  J.~Hisano, S.~Matsumoto, M.~Nagai, O.~Saito and M.~Senami,
  Phys.\ Lett.\  B {\bf 646}, 34 (2007)
  [arXiv:hep-ph/0610249].

\bibitem{MarchRussell:2008yu}
  J.~March-Russell, S.~M.~West, D.~Cumberbatch and D.~Hooper,
  JHEP {\bf 0807}, 058 (2008)
  [arXiv:0801.3440 [hep-ph]].

\bibitem{Yuan:2009bb}
  Q.~Yuan, X.~J.~Bi, J.~Liu, P.~F.~Yin, J.~Zhang and S.~H.~Zhu,
  JCAP {\bf 0912}, 011 (2009)
  [arXiv:0905.2736 [astro-ph.HE]].

\bibitem{Dent:2009bv}
  J.~B.~Dent, S.~Dutta and R.~J.~Scherrer,
  Phys.\ Lett.\  B {\bf 687}, 275 (2010)
  [arXiv:0909.4128 [astro-ph.CO]].

\bibitem{Zavala:2009mi}
  J.~Zavala, M.~Vogelsberger and S.~D.~M.~White,
  Phys.\ Rev.\  D {\bf 81}, 083502 (2010)
  [arXiv:0910.5221 [astro-ph.CO]].

\bibitem{Gondolo:1990dk}
  P.~Gondolo and G.~Gelmini,
  Nucl.\ Phys.\  B {\bf 360}, 145 (1991).

\bibitem{Kolb:1990vq}
  E.~W.~Kolb and M.~S.~Turner,
  Front.\ Phys.\  {\bf 69}, 1 (1990).

\bibitem{Holdom:1985ag}
  B.~Holdom,
  Phys.\ Lett.\  B {\bf 166}, 196 (1986).

\bibitem{Batell:2009yf}
  B.~Batell, M.~Pospelov and A.~Ritz,
  Phys.\ Rev.\  D {\bf 79}, 115008 (2009)
  [arXiv:0903.0363 [hep-ph]].

\bibitem{Bjorken:2009mm}
  J.~D.~Bjorken, R.~Essig, P.~Schuster and N.~Toro,
  Phys.\ Rev.\  D {\bf 80}, 075018 (2009)
  [arXiv:0906.0580 [hep-ph]].

\bibitem{Pospelov:2007mp}
  M.~Pospelov, A.~Ritz and M.~B.~Voloshin,
  Phys.\ Lett.\  B {\bf 662}, 53 (2008)
  [arXiv:0711.4866 [hep-ph]].

\bibitem{Feng:2008mu}
  J.~L.~Feng, H.~Tu and H.~B.~Yu,
  JCAP {\bf 0810}, 043 (2008)
  [arXiv:0808.2318 [hep-ph]].

\bibitem{Finkbeiner:2009mi}
  D.~P.~Finkbeiner, T.~R.~Slatyer, N.~Weiner and I.~Yavin,
  JCAP {\bf 0909}, 037 (2009)
  [arXiv:0903.1037 [hep-ph]].

\bibitem{Feng:2009mn}
  J.~L.~Feng, M.~Kaplinghat, H.~Tu and H.~B.~Yu,
  JCAP {\bf 0907}, 004 (2009)
  [arXiv:0905.3039 [hep-ph]].

\bibitem{Ackerman:2008gi}
  L.~Ackerman, M.~R.~Buckley, S.~M.~Carroll and M.~Kamionkowski,
  Phys.\ Rev.\  D {\bf 79}, 023519 (2009)
  [arXiv:0810.5126 [hep-ph]].

\bibitem{Buckley:2009in}
  M.~R.~Buckley and P.~J.~Fox,
  Phys.\ Rev.\  D {\bf 81}, 083522 (2010)
  [arXiv:0911.3898 [hep-ph]].

\bibitem{Ibe:2009mk}
  M.~Ibe and H.~B.~Yu,
  Phys.\ Lett.\  B {\bf 692}, 70 (2010)
  [arXiv:0912.5425 [hep-ph]].

\bibitem{Khrapak:2003}
S.~A.~Khrapak \etal,
Phys.\ Rev.\ Lett.\  {\bf 90}, 225002 (2003).

\bibitem{Khrapak:2004}
S.~A.~Khrapak \etal,
IEEE Transactions on Plasma Science {\bf 32}, 555 (2004).

\bibitem{Xue:2008se}
  X.~X.~Xue {\it et al.}  [SDSS Collaboration],
  Astrophys.\ J.\  {\bf 684}, 1143 (2008)
  [arXiv:0801.1232 [astro-ph]].
  
\bibitem{Abazajian:2010sq}
  K.~N.~Abazajian, P.~Agrawal, Z.~Chacko and C.~Kilic,
  arXiv:1002.3820 [astro-ph.HE].

\bibitem{Kamionkowski:1997xg}
  M.~Kamionkowski and A.~Kinkhabwala,
  Phys.\ Rev.\  D {\bf 57}, 3256 (1998)
  [arXiv:hep-ph/9710337].

\bibitem{Catena:2009mf}
  R.~Catena and P.~Ullio,
  JCAP {\bf 1008}, 004 (2010)
  [arXiv:0907.0018 [astro-ph.CO]].

\bibitem{Weber:2009pt}
  M.~Weber and W.~de Boer,
  arXiv:0910.4272 [astro-ph.CO].

\bibitem{Salucci:2010qr}
  P.~Salucci, F.~Nesti, G.~Gentile and C.~F.~Martins,
  arXiv:1003.3101 [astro-ph.GA].

\bibitem{Cline:2010ag}
  J.~M.~Cline, A.~C.~Vincent and W.~Xue,
  Phys.\ Rev.\  D {\bf 81}, 083512 (2010)
  [arXiv:1001.5399 [astro-ph.CO]].

\bibitem{Vogelsberger:2008qb}
  M.~Vogelsberger {\it et al.},
  MNRAS {\bf 395}, 797 (2009)
  [arXiv:0812.0362 [astro-ph]].

\bibitem{Kuhlen:2009is}
  M.~Kuhlen and D.~Malyshev,
  Phys.\ Rev.\  D {\bf 79}, 123517 (2009)
  [arXiv:0904.3378 [hep-ph]].

\bibitem{Hooper:2008kg}
  D.~Hooper, P.~Blasi and P.~D.~Serpico,
  JCAP {\bf 0901}, 025 (2009)
  [arXiv:0810.1527 [astro-ph]].

\bibitem{Yuksel:2008rf}
  H.~Yuksel, M.~D.~Kistler and T.~Stanev,
  Phys.\ Rev.\ Lett.\  {\bf 103}, 051101 (2009)
  [arXiv:0810.2784 [astro-ph]].

\bibitem{Profumo:2008ms}
  S.~Profumo,
  arXiv:0812.4457 [astro-ph].

\bibitem{Dado:2009ux}
  S.~Dado and A.~Dar,
  Mem.\ Sci.\ Astron.\ Ital.\ {\bf 81}, 132 (2010)
[arXiv:0903.0165 [astro-ph.HE]].

\bibitem{Biermann:2009qi}
  P.~L.~Biermann, J.~K.~Becker, A.~Meli, W.~Rhode, E.~S.~Seo and T.~Stanev,
  Phys.\ Rev.\ Lett.\  {\bf 103}, 061101 (2009)
  [arXiv:0903.4048 [astro-ph.HE]].

\bibitem{Katz:2009yd}
  B.~Katz, K.~Blum, J.~Morag, and E.~Waxman,
  MNRAS {\bf 405}, 1458 (2010)
  [arXiv:0907.1686 [astro-ph.HE]].

\bibitem{Grasso:2009ma}
  D.~Grasso {\it et al.}  [FERMI-LAT Collaboration],
  Astropart.\ Phys.\  {\bf 32}, 140 (2009)
  [arXiv:0905.0636 [astro-ph.HE]].

\bibitem{Galli:2009zc}
  S.~Galli, F.~Iocco, G.~Bertone and A.~Melchiorri,
  Phys.\ Rev.\  D {\bf 80}, 023505 (2009)
  [arXiv:0905.0003 [astro-ph.CO]].

\bibitem{Slatyer:2009yq}
  T.~R.~Slatyer, N.~Padmanabhan and D.~P.~Finkbeiner,
  Phys.\ Rev.\  D {\bf 80}, 043526 (2009)
  [arXiv:0906.1197 [astro-ph.CO]].

\bibitem{Barger:2009yt}
  V.~Barger, Y.~Gao, W.~Y.~Keung, D.~Marfatia and G.~Shaughnessy,
  Phys.\ Lett.\  B {\bf 678}, 283 (2009)
  [arXiv:0904.2001 [hep-ph]].

\bibitem{Falkowski:2010cm}
  A.~Falkowski, J.~T.~Ruderman, T.~Volansky and J.~Zupan,
  JHEP {\bf 1005}, 077 (2010)
  [arXiv:1002.2952 [hep-ph]].

\bibitem{MarchRussell:2008tu}
 J.~D.~March-Russell and S.~M.~West,
 Phys.\ Lett.\  B {\bf 676}, 133 (2009)
 [arXiv:0812.0559 [astro-ph]].

\bibitem{Shepherd:2009sa}
 W.~Shepherd, T.~M.~P.~Tait and G.~Zaharijas,
 Phys.\ Rev.\  D {\bf 79}, 055022 (2009)
 [arXiv:0901.2125 [hep-ph]].

\bibitem{TuckerSmith:2001hy}
  D.~Tucker-Smith and N.~Weiner,
  Phys.\ Rev.\  D {\bf 64}, 043502 (2001)
  [arXiv:hep-ph/0101138].

\bibitem{Chang:2008gd}
  S.~Chang, G.~D.~Kribs, D.~Tucker-Smith and N.~Weiner,
  Phys.\ Rev.\  D {\bf 79}, 043513 (2009)
  [arXiv:0807.2250 [hep-ph]].

\bibitem{Finkbeiner:2007kk}
  D.~P.~Finkbeiner and N.~Weiner,
  Phys.\ Rev.\  D {\bf 76}, 083519 (2007)
  [arXiv:astro-ph/0702587].

\bibitem{Cholis:2008vb}
  I.~Cholis, L.~Goodenough and N.~Weiner,
  Phys.\ Rev.\  D {\bf 79}, 123505 (2009)
  [arXiv:0802.2922 [astro-ph]].

\bibitem{Zurek:2008qg}
  K.~M.~Zurek,
  Phys.\ Rev.\  D {\bf 79}, 115002 (2009)
  [arXiv:0811.4429 [hep-ph]].

\bibitem{Katz:2009qq}
  A.~Katz and R.~Sundrum,
  JHEP {\bf 0906}, 003 (2009)
  [arXiv:0902.3271 [hep-ph]].

\bibitem{Cholis:2009va}
  I.~Cholis and N.~Weiner,
  arXiv:0911.4954 [astro-ph.HE].

\bibitem{Chen:2009ab}
  F.~Chen, J.~M.~Cline and A.~R.~Frey,
  Phys.\ Rev.\  D {\bf 80}, 083516 (2009)
  [arXiv:0907.4746 [hep-ph]].

\bibitem{Cirelli:2010nh}
  M.~Cirelli and J.~M.~Cline,
  Phys.\ Rev.\  D {\bf 82}, 023503 (2010)
  [arXiv:1005.1779 [hep-ph]].

\bibitem{Kane:2009if}
  G.~Kane, R.~Lu and S.~Watson,
  Phys.\ Lett.\  B {\bf 681}, 151 (2009)
  [arXiv:0906.4765 [astro-ph.HE]].

\bibitem{Zhang:2009dd}
 H.~Zhang, C.~S.~Li, Q.~H.~Cao and Z.~Li,
 arXiv:0910.2831 [hep-ph].

\bibitem{Kawasaki:1994sc}
  M.~Kawasaki and T.~Moroi,
  Astrophys.\ J.\  {\bf 452}, 506 (1995)
  [arXiv:astro-ph/9412055].

\bibitem{Cyburt:2002uv}
  R.~H.~Cyburt, J.~R.~Ellis, B.~D.~Fields and K.~A.~Olive,
  Phys.\ Rev.\  D {\bf 67}, 103521 (2003)
  [arXiv:astro-ph/0211258].

\bibitem{Kamionkowski:2008gj}
  M.~Kamionkowski and S.~Profumo,
  Phys.\ Rev.\ Lett.\  {\bf 101}, 261301 (2008)
  [arXiv:0810.3233 [astro-ph]].

\bibitem{Belikov:2009qx}
  A.~V.~Belikov and D.~Hooper,
  Phys.\ Rev.\  D {\bf 80}, 035007 (2009)
  [arXiv:0904.1210 [hep-ph]].

\bibitem{Profumo:2009uf}
  S.~Profumo and T.~E.~Jeltema,
  JCAP {\bf 0907}, 020 (2009)
  [arXiv:0906.0001 [astro-ph.CO]].

\bibitem{Huetsi:2009ex}
  G.~Huetsi, A.~Hektor and M.~Raidal,
  Astron.\ Astrophys.\  {\bf 505}, 999 (2009)
  [arXiv:0906.4550 [astro-ph.CO]].

\bibitem{Cirelli:2009bb}
  M.~Cirelli, F.~Iocco and P.~Panci,
  JCAP {\bf 0910}, 009 (2009)
  [arXiv:0907.0719 [astro-ph.CO]].

\bibitem{Kanzaki:2009hf}
  T.~Kanzaki, M.~Kawasaki and K.~Nakayama,
  Prog.\ Theor.\ Phys.\  {\bf 123}, 853 (2010)
  [arXiv:0907.3985 [astro-ph.CO]].

\bibitem{Crocker:2010gy}
  R.~M.~Crocker, N.~F.~Bell, C.~Balazs and D.~I.~Jones,
  Phys.\ Rev.\  D {\bf 81}, 063516 (2010)
  [arXiv:1002.0229 [hep-ph]].

\bibitem{Calvez:2010jq}
  A.~Calvez, W.~Essey, M.~Fairbairn, A.~Kusenko and M.~Loewenstein,
  arXiv:1003.1113 [astro-ph.HE].

\bibitem{PalomaresRuiz:2010pn}
  S.~Palomares-Ruiz and J.~M.~Siegal-Gaskins,
  JCAP {\bf 1007}, 023 (2010)
  [arXiv:1003.1142 [astro-ph.CO]].

\end{thebibliography}
\end{document}